# A 3D porous MXene/PNIPAAm hydrogel composite with advanced degradation stability and control of electronic properties in air

*Sitao Wang[1], Chen Jiao[2], Gerald Gerlach[1], Julia Körner[3,]\**

[1]Institute of Solid-State Electronics, Dresden University of Technology, 01062 Dresden, Germany

[2]Leibniz Institute of Polymer Research Dresden, 01069 Dresden, Germany

[3]Department of Electrical Engineering and Computer Science, Leibniz University Hannover, 30167 Hannover, Germany

\*Correspondence: koerner@geml.uni-hannover.de

ABSTRACT

This study reports the fundamental investigation of a novel composite material consisting of MXene ($Ti_3C_2T_x$) and a stimulus-responsive hydrogel (Poly(*N*-isopropylacrylamide) – PNIPAAm). In contrast to previously reported integration of MXene and hydrogels, the material described here offers a tunable porous structure for volatile organic compound (VOC) sensing that maintains its stability and responsiveness in any kind of gaseous environment and in a dried state. Furthermore, its synthesis is significantly simpler and more efficient compared to other studies on porous MXene composites.

The presented study focuses on a fundamental investigation of the synthesized MXene/PNIPAAm composite's properties. Thereby, the fabrication and comparison of pure MXene and composite samples featuring either a compact or a highly porous three-dimensional microstructure, reveal unique properties with respect to: (i) controllable three-dimensional spatial arrangement of MXene instead of the prevalent stacked-sheet structure, (ii) reduction of oxidation-induced degradation of MXene and substantially enhanced stability over the course of three months, and (iii) tunable electronic states in response to gas interactions. Material characterization is conducted by scanning electron microscopy and rheology to assess the microstructural and mechanical properties, and in a chemiresistive measurement setup for determination of electrical properties and the evaluation of the composite's potential for VOC sensing in a gaseous environment with the test analyte acetone. These investigations reveal fundamentally novel material effects and properties that address some of the key MXene-related challenges. Additionally, the interplay between the MXene and the hydrogel enables unprecedented opportunities for enhancing the sensing potential of stimulus-responsive hydrogels, specifically in gaseous environments.

## 1. Introduction

MXenes are a class of two-dimensional transition-metal carbides/nitrides/carbonitrides that has gained substantial attention due to its versatility of composition and tuneability of properties, as well as many different potential applications, such as in energy storage and conversion, electronics or biomedical contexts.[1-3] However, despite extensive research efforts, fundamental challenges remain, which have been listed and ranked by experts in the field in 2021 as 32 major points to be addressed in the next decade.[4] Among the top priorities are an improvement of chemical and temperature stability, the development of methods for creating fully three-dimensional nanostructures with controlled alignment or orientation of MXene flakes, and the fundamental understanding and control of material properties such as electronic, optical and magnetic.[4]

Reported approaches of addressing the degradation issue include defect passivation, chemical intercalation,[5,6] and, particularly for aqueous media, the combination with a polymer material.[7] The latter can furthermore result in altered and enhanced material properties, which in turn is related to the exploration of new use cases. For example, the fabrication of MXene/polydopamine hybrid films with enhanced mechanical properties and ambient stability through the self-polymerization of dopamine monomers on MXene surfaces has been reported.[8] The uniform atomic-level thickness of the polymer enables MXene to be structured in an idealized parallel manner, effectively preventing the penetration of oxygen and moisture. In another work, a conductive MXene/gelatin composite with excellent environmental stability and self-adhesiveness for multifunctional sensor applications was demonstrated. The research findings suggested that gelatin molecules can be easily adsorbed onto the MXene sheets, forming a protective shield that prevents restacking and oxidation, resulting in a highly stable MXene/gelatin dispersion, as opposed to MXene aqueous dispersion.[9]

In all of these cases, the polymer material is attached to or fabricated on the already formed MXene structure and assumes the role of a passivation layer.

However, the combination of MXene and polymers can also be directed towards composite materials where the diverse properties of the MXene are employed to alter and engineer the polymer behavior and, on the other hand, polymer functionalities benefit the MXene component.

In general, the creation of polymeric nanohybrids by incorporation of inorganic materials such as metal nanoparticles, metal oxides, graphene, or carbon-based nanostructures,[10-15] has been reported to lead to synergistic effects between the different phases in such a composite. The polymer matrix provides a high surface area with controllable and tunable pore size and distribution as well as a scaffold for distribution and arrangement of the additives, while the non-polymeric components offer advantages such as electrical conductivity and versatile surface chemistry.[16-18]

It has already been demonstrated that the specific material composition and (surface) chemistry of MXenes and respective composites provide a promising basis for applications in the context of gas sensing and volatile organic compound (VOC) detection.[19-22] For this purpose, the material needs to be specifically tailored to reach high sensitivities, for example by partial oxidation and spatial arrangement of the MXene sheets.[23-25]

Another material candidate for sensing and actuation applications, in particular in biomedical contexts, are stimulus-responsive (smart) hydrogels.[26,27] This is a specific class of polymers whose key feature is a reversible volume change as a reaction to external stimuli such as light,[28,29] pH,[30] temperature,[31] ionic strength,[32] and biological molecules.[33]

However, the volume change of smart hydrogels is mainly mediated by the absorption and release of liquid, which has largely limited the application scope to aqueous environments and humidity sensing.[34,35] Only recently has it been explored how this boundary condition can be overcome to enable harnessing of the selective and sensitive volumetric response for volatile organic compound (VOC) detection in gaseous environments.[36-39] In prior work, we have successfully demonstrated a novel approach to specifically tailor the structure of polymer networks for long-term stable porosity in air with largely varying humidity (5% to 100%), and the applicability for the detection of the organic test analytes acetone and isopropanol. The specific smart hydrogel suitable for the chosen test gases was found to be Poly(*N*-isopropylacrylamide) (PNIPAAm), but the developed porosity engineering techniques are of general nature and applicable for polymer materials in general.[38,39]

Please note that PNIPAAm is in principle a temperature-responsive hydrogel that undergoes a volume-phase transition in liquid environments at a lower critical solution temperature (LCST) of 33 °C.[40,41] However, this responsiveness is not utilized for the use of VOC detection in air, where a change of the hydrophilic-hydrophobic balance in the polymer network is only expected to have a minimal effect due to the absence of a solution environment. Instead, the functional groups of the polymer directly interact with the VOC and a detailed explanation of this mechanism is presented in prior work.[38] Furthermore, all investigations reported in the following have been conducted under cleanroom conditions under a controlled temperature of (20 – 22) °C, which is significantly below the LCST of PNIPAAm hydrogel.

While the potential of PNIPAAm for gas sensing has been demonstrated, the aforementioned mechanism of direct interaction between the polymer's functional groups and the VOC poses challenges with regard to (i) limited sensitivity and swelling response due to the adsorption and desorption dynamics of the organic compounds with the polymer, and (ii) the reliable transduction of the smart hydrogel's volume change response.

In both aspects, MXene is an ideal material to be combined with smart hydrogels to (i) enhance the responsiveness for VOCs, and (ii) equip the intrinsically insulating polymer with electrical conductivity to extend the array of potential transduction concepts in the sensing context. To date, the majority of reported work on polymer-MXene composites is focusing on either one of the components as the active material, while the other one is used as a support, e.g., scaffold, encapsulation, or conductivity enhancer. An overview is provided in Table 1. For PNIPAAm hydrogel, no combination with MXene for gas sensing, in particular where the responsiveness of both materials is considered, has been reported to date.

**Table 1** Overview of reported studies on MXene-polymer composites and their applications.

| Material | Application | Ref. |
|---|---|---|
| MXene on 3D polymer framework | VOC gas sensing<br>Sensing material: MXene; polymer (PVA/PEI) as 3D scaffold | [42] |
| MXene/polyurethane core-sheath fibers | VOC gas sensing<br>MXene: VOC sensing + conductivity; polyurethane: stretchability | [43] |
| MXene/PNIPAAm composite | Soft manipulators and strain sensors<br>Hydrogel: softness; MXene: conductivity + photothermal response enhancer for hydrogel actuation | [44] |
| Cyclodextrin-encapsulated MXene/PNIPAAm composite | Absorbent for phenols in waste water<br>Both combined: enhanced binding of target; cyclodextrin as 3rd component to extent material lifetime | [41] |
| MXene/PNIPAAm composite | Smart compression sensor (temperature + stress)<br>Hydrogel: softness + stretchability; MXene: conductivity | [45] |
| MXene nanocomposite polymers | Conductive 3D printable polymer for soft electronics<br>MXene: conductivity; polymer: printable resin | [46] |

In the work presented here, we have created a novel composite material comprising the smart hydrogel Poly(*N*-isopropylacrylamide) (PNIPAAm) and MXene ($Ti_3C_2T_x$) that for the first time enables the exploration of the phenomenon of a mutual beneficial interaction and combination of properties towards new functionalities, with an exemplary application in gas sensing. It relies on a recently developed process of structural porosity engineering of polymers[38,39] to achieve an interplay between both components.

Thereby, the MXene on the one hand alters the mechanical and electrical properties of the hydrogel, also with respect to VOC sensing mechanisms. On the other hand, the interplay with the polymer enables the necessary MXene modifications beneficial for gas sensing (partial oxidation, three-dimensional spatial arrangement) mentioned above and, furthermore, results in unprecedented opportunities for addressing some of the challenges of MXenes mentioned in the beginning. The unique features of the composite material include (i) the spatial arrangement of MXene flakes in a fully three-dimensional structure instead of the usual stacked-sheet configuration, (ii) control over the electronic response to a gas interaction on the range from metallic to semiconducting by tailoring the composite's porosity, and (iii) a reduction of oxidation-induced degradation of MXene and unprecedented stability over the course of three months.

While integrating MXene with hydrogel materials is not a novel concept, the presented dried composite sample stands out by offering a tunable porous structure for gas sensing that maintains its stability in any kind of humid VOC environment, a feature that has not been documented in the current literature. Additionally, the fabrication process is significantly simpler and more efficient compared to other studies on porous MXene composites.

In the following, the fabrication and characterization of pure MXene and MXene/PNIPAAm composite samples is described with respect to the fundamental material properties and its potential for VOC sensing. Besides scanning electron microscopy and rheology for assessing microstructure and mechanical properties, chemiresistive measurements are employed to gain insights into the electronic behavior of the samples in response to an exposure to acetone as an exemplary test VOC. Chemiresistive sensing utilizes the change of electrical impedance of an electrode structure (usually interdigitated electrodes – IDEs) containing the sensing material as a dielectric medium. The electrical properties of the sample material are affected by the interaction with an analyte (or other stimulus in case of a hydrogel), which in turn is detected and monitored by the IDE's impedance.

Comprehensive analysis of the response to the exemplary organic compound acetone enables the assessment of the electronic states as well as the degradation stability with regard to oxidation of the MXene phase of the MXene/PNIPAAm composite. The comparison with pure but three-dimensionally arranged MXene furthermore reveals the role of the hydrogel and the interplay between the two components, indicating a novel path towards adjustable and controlled electronic properties and a tunable spatial MXene arrangement.

## 2. Materials and methods

**2.1 Studied Materials**

To assess the performance of individual components, pure MXene ($Ti_3C_2T_x$) and MXene/PNIPAAm composite materials were synthesized as bulk (1 cm × 0.5 cm × 0.2 cm), and as thin film samples on interdigitated electrodes (IDEs). Bulk samples were fabricated with molds (glass slides of 76 mm × 52 mm with a 500 μm thick Teflon spacer) and the samples cut into smaller pieces of approximately 0.001 mg and 0.004 mg for pure MXene and the composite, respectively, for analysis. The thin films were created by drop-casting of 4 μL of precursor solution (pure MXene) or by molding (composite samples) on the desired target substrate (Figure S1, Supporting Information). The IDEs in the study comprise of either a gold or platinum metallization layer (same electrode pattern) and different base materials (polyimide, glass, ceramics) to match the material properties of the attached material (either pure MXene or the composite), ensure good adhesion and therefore a reliable electrical impedance read-out. Table 2 provides an overview of the studied sample types and used characterization methods since not all techniques have been applied to every sample type due to feasibility. All investigations have been performed for at least three samples of each type to ensure reproducibility and reliability of the reported findings. Please refer to the Supporting Information for details about synthesis and characterization.

**Table 2** Overview of studied samples and applied characterization technique for different sample types. Used abbreviations: FD - freeze-dried; AD - air-dried; PEG - polyethylene-glycol; OM - optical microscopy; SEM - scanning electron microscopy; EDX - energy-dispersive X-ray spectroscopy; Rheo - rheology; CR - chemiresistive (short-term, i.e., as-fabricated and long-term analysis with repeated testing over weeks).

| Category | Specific properties | OM | SEM | EDX | Rheo | CR short | CR long |
|---|---|---|---|---|---|---|---|
| **Pure Mxene** | Bulk, FD | ✓ | ✓ | ✓ | | | |
| | On-chip, AD | ✓ | ✓ | ✓ | | ✓ | |
| | On-chip, FD | ✓ | ✓ | ✓ | | ✓ | ✓ |
| **MXene/PNIPAAm composite** | Bulk, FD, with PEG | ✓ | ✓ | ✓ | ✓ | | |
| | Bulk, FD, without PEG | ✓ | ✓ | ✓ | | | |
| | On-chip, AD, with PEG | ✓ | ✓ | ✓ | | ✓ | |
| | On-chip, FD, with PEG | ✓ | ✓ | ✓ | | ✓ | ✓ |

**2.2 Sample fabrication**

In all cases where MXene was used, a corresponding suspension of a concentration of 30 mg/mL was first made from powder and then used for further processing. This concentration was chosen based on prior experiments, which showed that 30 mg/mL is the highest MXene content that still enables a successful polymerization in combination with the target hydrogel PNIPAAm. Higher MXene contents cause incomplete polymerization and subsequent disintegration of the samples. Lower MXene concentrations are studied in the context of mechanical properties of the composite as described in section 3.2.5.

The 3D pure MXene network was obtained by freezing-induced preassembly of the MXene suspension followed by nanosheets cross-linking in hydrochloric acid (HCl). The former offers a tunable microstructure, where the formation of ice crystals plays a crucial role in determining the final structure.[47,48] Therefore, two different freezing temperatures (-196 °C and -20 °C) were studied. The latter ensures a reinforced scaffold by intercalating cations between the sheets at a high concentration, resulting in three-dimensional interlinking of 2D MXene.[49] The 3D MXene structures were fabricated on IDEs with platinum metallization by drop-casting.

The MXene/PNIPAAm composite was created by mixing the monomer precursor with the prefabricated MXene suspension. Additionally, long-chain poly(ethylene glycol) (PEG) was included as a pore-forming agent as described in previous work.[39] It creates intrinsic voids during polymerization, resulting in a large surface area of the polymer after PEG removal. The quantity of the applied porogen must be sufficient to create adequate voids in the matrix without exceeding a threshold that disrupts the cross-linking of the entire sample, potentially causing polymerization failure. To align the mechanical properties of the material with the IDE substrate and ensure good adhesion between them, IDEs made of platinum on a flexible polyimide substrate and gold on a ceramic substrate were utilized.

To stabilize the porous microstructure of the fabricated materials for the intended use of gas sensing, post-fabrication processing includes the steps of washing (to remove all unreacted chemicals), freeze-drying in liquid nitrogen and, only in case of the hydrogel-containing samples, subsequent conditioning in high humidity to allow the polymer chains to reconfigure to an equilibrium state. This procedure has been previously developed and demonstrated to enable a stable porous microstructure of PNIPAAm hydrogel in gaseous environments of varying relative humidity and for VOC sensing.[39] Due to the reduced sample size and the improved mechanical properties of the MXene/PNIPAAm composite, in the presented experiments, the first test in the organic/humid gas atmosphere was used to stabilize the material structure, rather than conditioning the sample in high humidity for 3 days. This approach suffices to ensure reproducible results in the subsequent testing within the same gaseous environment.

## 2.3 Material characterization

Optical and scanning electron microscopy (SEM) are employed to examine the appearance and microstructures of the fabricated materials. For SEM analysis, the composite samples are sputter-coated with 10 nm of gold to prevent charging effects due to the polymer's low electrical conductivity. No coating is required for the pure MXene material. Elemental distribution within the samples is determined using energy-dispersive X-ray spectroscopy (EDX).

The mechanical properties of the bulk composite samples are assessed through rheology. Rheological frequency sweep tests are carried out after hydrogel curing to determine how MXene additives affect the hydrogel's storage ($G'$) and loss moduli ($G''$) over a defined frequency range from 0.1 to 100 rad/s (ARES-G2, TA Instruments, USA).

To study the chemiresistive performance of pure MXene and the MXene/PNIPAAm composite, IDEs equipped with the respective material are placed in a sealed chamber where desired environmental conditions can be created and the IDE impedance is measured with a digital multimeter (Fluke 45 Dual Display Multimeter, Germany). In the presented studies, acetone is used as the exemplary test analyte to enable a comparison with prior work.[38,39] Furthermore, this setup is also employed to assess the degradation stability of the investigated samples.

Figure 1 depicts the general setup. For IDEs equipped with pure MXene, the chamber is flushed with dried nitrogen in between liquid acetone injections with a syringe pump (Figure 1a). For the MXene/PNIPAAm composite samples, the atmosphere needs to be humid to prevent the hydrogel from drying out and to create a measurable resistance signal. With a dried nitrogen background as in the case of pure MXene, no output signal could be obtained.

Predefined humidity conditions are created by placing two beakers of 1 mL of deionized water in the sealed chamber which leads to the presence of excess water molecules and, therefore, a condition of constant 100 % relative humidity. Liquid acetone is then added to this state by the syringe to create the desired organic solvent atmosphere. After a predefined exposure time interval, the solvent is removed by purging the chamber with a high humidity air flow created by bubbling dried nitrogen through a bottle of DI water with a flow rate of 1.8 L/min (Figure 1b). To ensure that the sample response is not influenced by the two different ways of creating a humidity environment (static and dynamic flow), control measurements have been performed with a commercial humidity sensor (Hytelog-RS232, B + B sensors, Germany). The results depicted in Figure 1c clearly indicate the stability of the created environment, regardless of the method used. Further details about the experimental setup, the calculation of injected liquid organic solvent volume and corresponding concentrations are outlined in detail in Figure S2 (Supporting Information) and our previous work.[38,39]

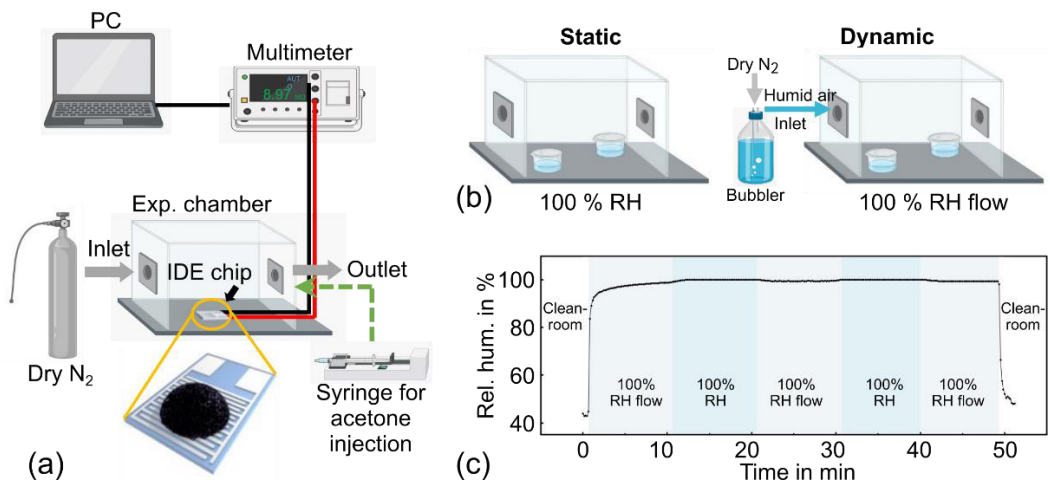

**Figure 1** Setup for IDE characterization: a) gas sensing test, b) setup modification for creation of a static saturated water vapor atmosphere (left) and dynamic humid air flow (right), c) baseline measurement of the relative humidity in the chamber created by either static conditions or dynamic humidity flow (see subfigure b).

## 3. Results

### 3.1 Analysis of pure MXene

In the following, the results of the pure MXene samples (bulk and on IDE) are described to establish a base line for comparison with the composite. Figure 2 depicts the fabrication process, SEM images and IDE output signals for the respective samples.

Please note that the fabrication steps in Figure 2a are shown for the bulk sample but were carried out in a similar way for the fabrication on IDE. In this case, instead of applying the MXene suspension to a mold, it was drop-casted on the IDE.

*3.1.1 Microstructure*

It needs to be noted here that the freezing temperature during the ice-templating process (-20 °C exemplarily indicated in Figure 2a) can significantly influence the appearance and mechanical stability of the resulting MXene structure (Figure S3, Supporting Information). Samples created by freezing in liquid nitrogen (-196 °C) resulted in a fragile scaffold, with the material falling apart. In addition, during quenching, the large temperature difference led to rapid nucleation and directional growth of ice crystals, leaving behind channel-like voids across the cross-section, with an orientation depending on the applied temperature gradient.[50] In contrast, when freezing at a higher temperature (-20 °C), the slow cooling rate allows ice crystals to develop omnidirectionally, leaving uniform cellular structures after drying.[51] Furthermore, this also promotes the formation of integral sections within the highly porous network, which are stiffer compared to the heterogeneous structures containing one-directional channel-like pores achieved at -196°C. Therefore, a freezing temperature of -20°C has been selected in this study for sensing applications. In SEM, the bulk pure MXene structure frozen at -20 °C exhibits a honeycomb-like porous network with a dark black color and shiny metallic features on the surface as shown in Figure 2b. The appearance under the optical microscope shows a uniformly gray and shiny coloring.

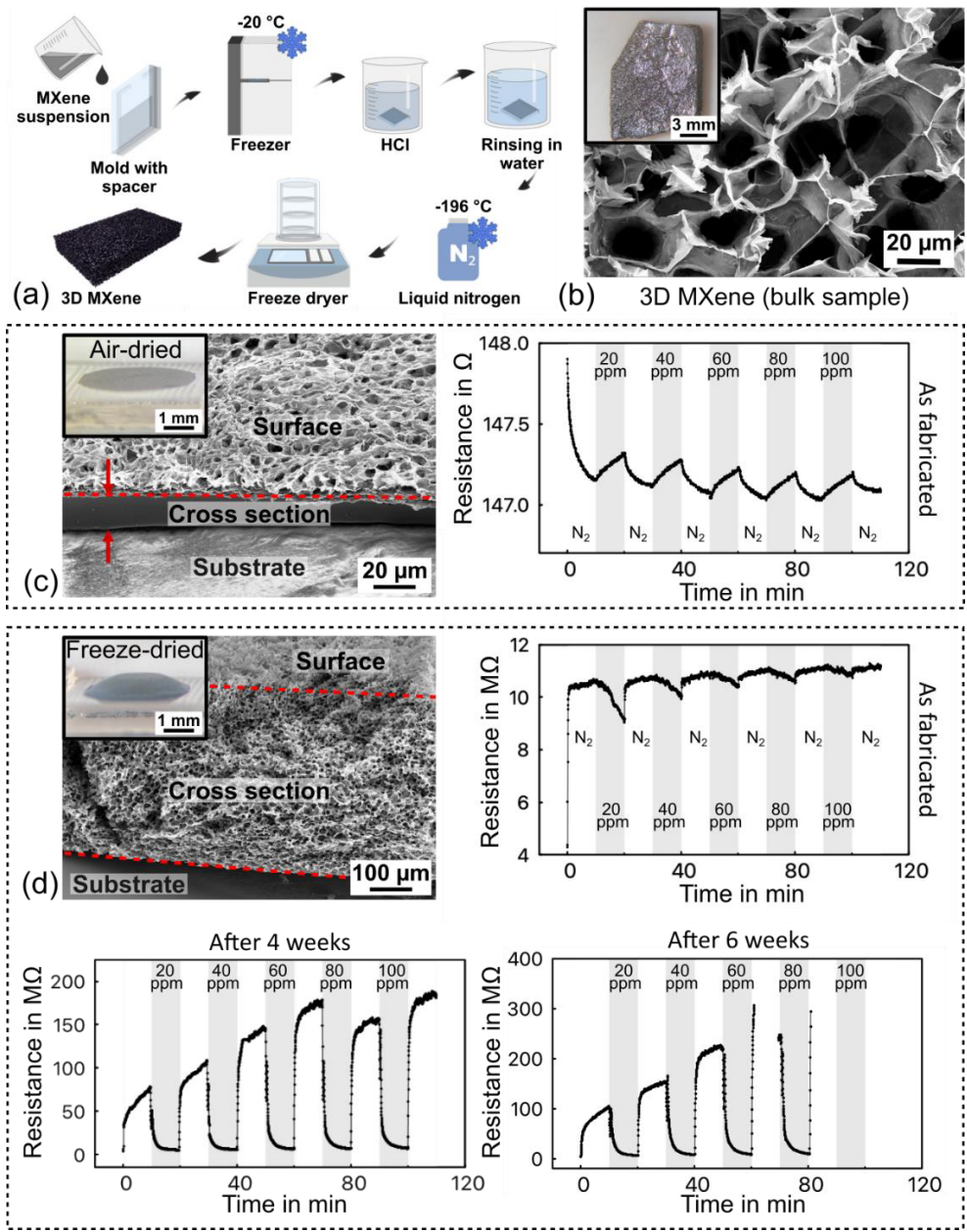

**Figure 2** 3D bulk pure MXene: a) fabrication process, b) optical microscope (inset) and SEM images of the obtained material. c, d) Optical microscope (insets) and SEM images of samples on IDEs (fabricated by drop casting) and the respective results of the chemiresistive measurements in (20 – 100) ppm gaseous acetone with a dry nitrogen background: c) air-dried compact MXene and d) freeze-dried 3D porous MXene. The 3D sample was repeatedly measured over the course of six weeks. IDE measurements were performed on as-fabricated samples without prolonged storage. Both samples were fabricated using the same volume of MXene suspension and a concentration of 30 mg/mL.

Figures 2c,d depict the pure MXene on IDEs that have been treated by two different procedures after fabrication and rinsing. In one case (Figure 2c), the sample was left to air-dry under cleanroom conditions (22 °C, 45 % RH), while in the second case, freeze-drying at -196 °C was applied (Figure 2d). This comparison was done to clarify if the pure MXene responds to different drying procedures in the same way as the PNIPAAm hydrogel (see[39]).

In accordance with the previous investigations of hydrogel, air-drying of the MXene results in an irreversible collapse of the porous network. Despite having a rugged surface morphology with recognizable pores, the capillary forces generated during water evaporation lead to the stacking of material layers with a reduced thickness of several microns and a densely packed internal structure.[52] In contrast, the 3D MXene network is preserved through freeze-drying at -196 °C in the same way that has been established for hydrogel material in previous studies.[39] The resulting MXene layer thickness amounts to several 100 μm, with the same amount of solution being used as in the air-dried case. The SEM image shows a homogeneous pore size and distribution with honeycomb-shaped voids along the cross-section (Figure 2d).

*3.1.2 IDE impedance measurements*

Chemiresistive measurements were carried out in an acetone atmosphere with concentrations ranging from 20 to 100 ppm (by adding of a corresponding amount of liquid solvent to the chamber as described above) and using dry nitrogen as the background and purge gas. The results for air- and freeze-dried IDE samples are also shown in Figures 2c, d. The dense MXene layer obtained by air-drying exhibits a resistance increase when exposed to acetone environment, consistent with findings from other published work.[53,54] However, the values show almost no dependence on the acetone concentration. The porous 3D MXene sample exhibits a reversed response, i.e., a resistance decrease, and a dependence on the acetone concentration. The resistance drop became more pronounced in the second test that was conducted four weeks after the initial test and subsequent storage under ambient cleanroom conditions, demonstrating a clearer correlation: with increasing acetone concentration, a larger resistance decrease occurred. However, the sample became immeasurable in the third test after six weeks of storage, as the resistance value increased to a very high level.

EDX analysis was performed on the 3D (freeze-dried) pure MXene sample right after fabrication and after exposure to acetone atmosphere for three tests to discern any differences. As depicted in Figure 3, the oxygen content of the as-fabricated MXene was notably lower than that of the sample after tests in acetone, indicating oxidation of the material.

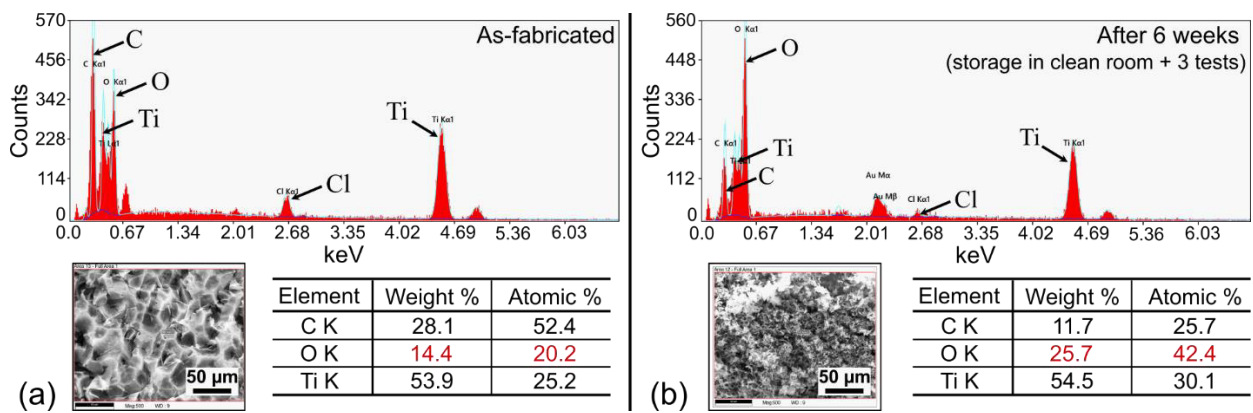

**Figure 3** EDX analysis results of 3D (freeze-dried) MXene samples: a) as-fabricated, b) after three tests in acetone atmosphere over a period of six weeks. In between impedance measurements, the sample was stored under cleanroom conditions.

## 3.2 Analysis of the MXene/PNIPAAm composite

The MXene/PNIPAAm composite with integrated PEG was fabricated as depicted in Figure 4a. The porogen PEG is necessary to create a porous hydrogel structure and, in contrast to ice crystal formation, it enables a more controlled approach as found in previous studies.[39] In contrast to the pure MXene, no freezing and HCl treatment are required as the polymer provides the scaffold and linkage for the MXene flakes.

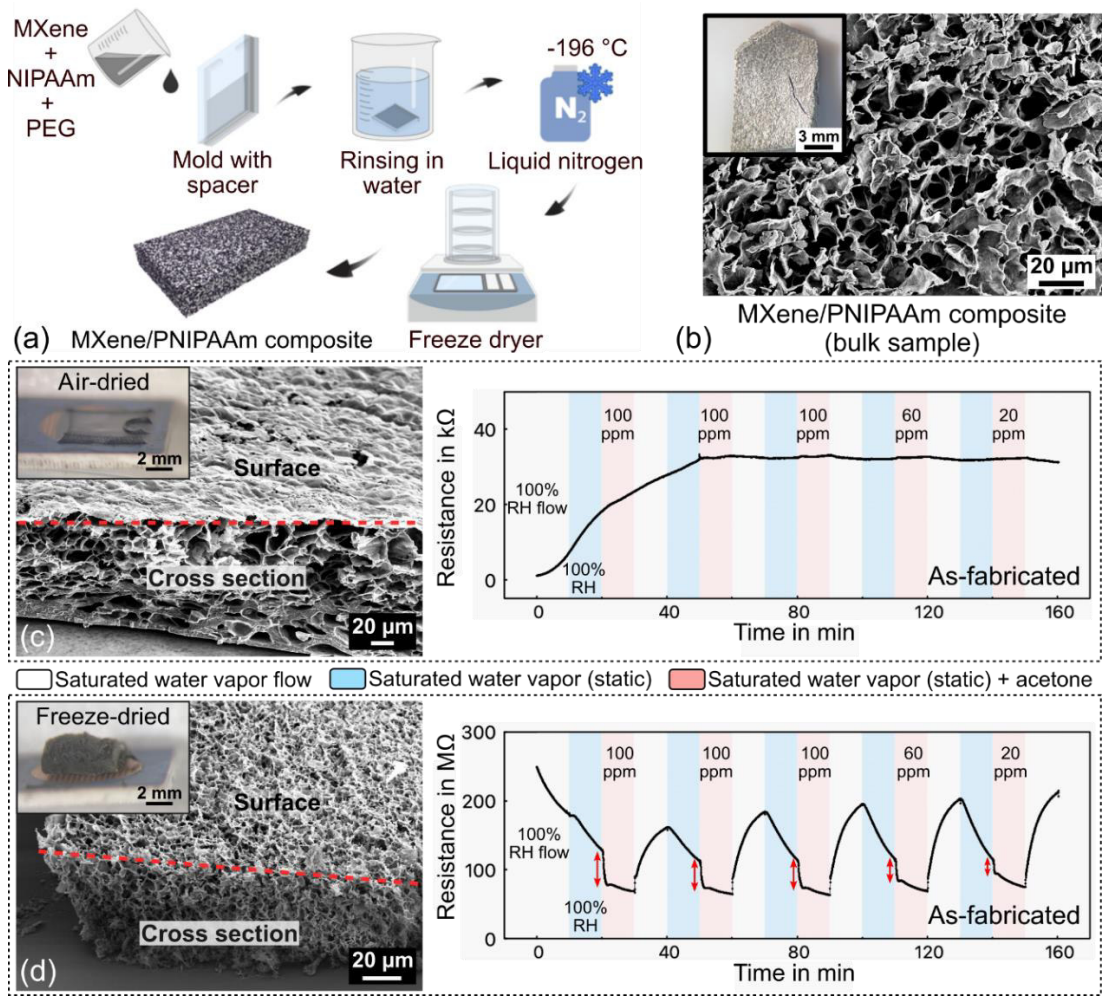

**Figure 4** Porous MXene/PNIPAAm composite: a) Fabrication process, b) bulk sample images (inset: optical microscope; main: SEM). Due to the synthesis process, the MXene is fully integrated with the polymer matrix and consequently, no individual MXene sheets can be identified. The homogeneous MXene distribution is instead evidenced by the EDX-mapping of the titanium element depicted in Figure 5. c, d): Optical microscope (inset), SEM images and chemiresistive analysis of MXene/PNIPAAm composite samples on IDE (fabricated by molding): c) air-dried compact and d) freeze-dried porous composite in response to gaseous acetone with a high humidity background. IDE measurements were performed on as-fabricated samples without prolonged storage. Both samples were fabricated with the same amount of precursor solution and a MXene concentration of 30 mg/mL.

*3.2.1 Microstructure*

In contrast to the almost completely grayish/black color of the pure MXene sample (Figure 2b), the MXene/PNIPAAm composite shows a black and white appearance under optical microscope as depicted in Figure 4b, indicating the presence of the polymer within the network. Furthermore, instead of the semi-transparent thin walls of pure MXene structures, the composite's morphology is more akin to the plain polymer, with embedded MXene flakes integrated into the polymeric matrix.[39] For reference, optical microscope and SEM images of pure PNIPAAm hydrogel featuring a clear white color are depicted in Figure S4 in the Supporting Information.

The determination of the composite's pore size is challenging due to formation of an overlapping irregular network of the polymer walls with the embedded MXene nanosheets. This effect is attributed to the relatively high concentration of the used MXene suspension. Overall, the pore diameter in the composite is estimated to be approximately 5 µm, which is in the same range but a bit smaller than the pure 3D MXene. Achieving similar pore dimensions would require advanced techniques for controlling ice growth and the utilization of porogens. Since the focus of the presented work is on the fundamental analysis of the novel composite's properties, a comprehensive investigation of the controlled porosity is beyond the scope of this work.

The bulk sample has been freeze-dried for microstructural analysis. For the on-chip samples, two different drying methods (air- and freeze-drying) have been employed to study the influence on the sample morphology and consequently the gas sensing performance. By comparing the SEM images in Figures 4c,d it is clearly evident that freeze-drying results in a finer porous structure, while air-drying leads to larger pores that are partially deformed and collapsed due to the slow evaporation of water molecules during drying. The most notable difference occurs on the sample surface: the freeze-dried material is equally porous on the surface and on the inside with interconnected pores throughout its entirety. In contrast, the surface of the air-dried sample forms a more skin-like closed layer that separates the inner porous network from the surroundings. These findings are consistent with previous studies of PNIPAAm.[39] Further detailed SEM images as well as a comparison to the pure MXene samples can be found in the Supporting Information (Figure S5).

With regard to the overall sample appearance (inset optical microscope images in Figures 4c,d, air-drying results in a very thin layer on the IDE for air-drying, with some material accumulating around the edges. This accumulation can be attributed to the coffee ring effect that has been observed for drying of polymer droplets containing solid-phase colloidal particles.[55] In contrast, freeze-drying preserves the structure and thickness of the original water-containing hydrogel. Please note that the same amount of precursor solution was used for both samples, resulting in a comparable watery sample thickness directly after synthesis.

*3.2.2 EDX analysis*

Since SEM images provide only surface-morphological information with embedded MXene sheets remaining mostly invisible, EDX was used to confirm the distribution of MXene within the polymer matrix. The results depicted in Figure 5 verify the homogeneous spatial distribution of MXene within the matrix as evidenced by the titanium element. Furthermore, the EDX mapping is shown for the as-fabricated composite as well as for a sample after 15 months of repeated cycling in acetone and humid atmosphere with prolonged storage under ambient cleanroom conditions in between. The extracted weight and atomic percentage values of the relevant elements carbon, oxygen and titanium indicate a substantial oxidation of the material since the amount of oxygen has doubled while the other two elements remain almost constant. This is similar to the results of the pure MXene (Figure 3) but in that case, the strong oxidation was already evident after six weeks with the material becoming unresponsive in the repeated chemiresistive testing (Figure 2). In contrast, the composite remained functional for at least three months as shown in Figure 6.

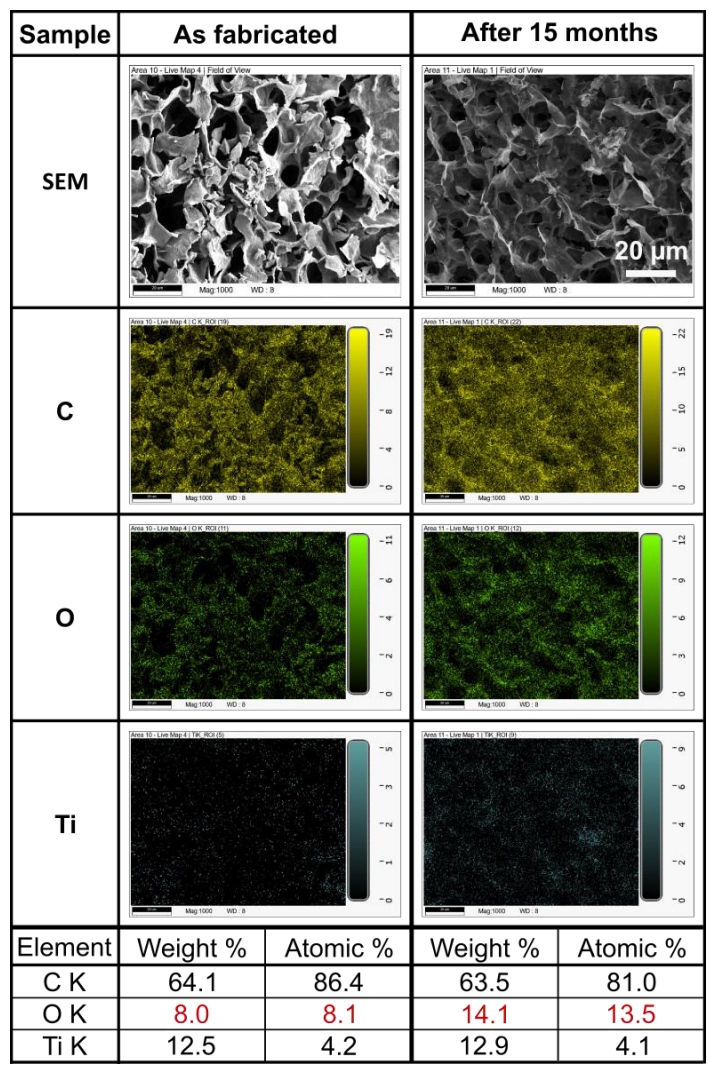

**Figure 5** Energy-dispersive X-ray spectroscopy (EDX) mapping results of an as-fabricated bulk MXene/PNIPAAm composite sample (left column), and an on-chip composite sample after 15 months of repeated testing in acetone with high humidity and storage under ambient cleanroom conditions in between. The scale bar applies to all images. Extracted weight percentage and atomic percentage values for the relevant elements carbon (C), oxygen (O) and titanium (Ti) are summarized in the table.

*3.2.3 Chemiresistive IDE analysis*

Figures 4c,d depict optical microscope and SEM images of the fabricated composite material on IDEs and the corresponding chemiresistive measurements in varying environmental conditions.

As in the case of pure MXene, air-dried and freeze-dried samples have been studied. It is clearly evident that the drying method exerts a significant influence on the sample structure as described in the microstructural comparison above. This significant difference in structure is further reflected in the gas sensing performance of the composite material. As depicted in Figure 4c, the limited surface area of the air-dried composite results in a constant resistance with no response to additional acetone molecules. An initial increase in resistance is observed up to approximately 50 mins and then stabilizes upon reaching equilibrium.

In contrast, the freeze-dried MXene/PNIPAAm composite exhibits a distinct response depending on the gaseous environment and it furthermore shows an overall significantly increased resistance value of about 100 MΩ compared to the air-dried counterpart (Figure 4d). This can likely be attributed to the highly porous structure, which leads to a larger spacing in between the embedded MXene sheets.

With the introduction of saturated water vapor (first as dynamic flow to clean the chamber and then in the static environment with the flow turned off), a decrease in resistance occurs due to the interaction of water molecules with the sample. Upon exposure to gaseous acetone, a sharp drop in resistance is observed (indicated by red arrows), followed by a further decrease but with much smaller slope.

When the chamber is then purged with water vapor only, the resistance immediately starts to increase again in the vapor flow and decreases in the static water vapor condition (flow turned off). This curve shape is consistently observed but the magnitude of the initial drop and the subsequent downward slope due to the introduction of acetone are dependent on the acetone concentration. A hypothesis for explaining these observations in the response to acetone is provided in the discussion section 4.3 below.

Overall, the magnitude of the initial resistance drop and the response time, i.e., the slope of the curve, are dependent on the acetone concentration. A higher concentration results in a larger resistance change and longer response time. The corresponding response times $t_{90}$ and magnitudes of the initial resistance change for the acetone response of the freeze-dried composite are listed in Table S1 in the Supporting Information. When comparing the sharp drop induced by exposure to an acetone atmosphere with existing research on polymer-based composite materials for acetone sensing, the calculated response times are in a very similar range.[42,56]

Another observation that is made from the first three cycles of the same acetone concentration (100 ppm) is a settling effect of the composite material. This is attributed to the almost inevitable conditioning process characteristic for hydrogel sensing materials after preparation or introduction into a new environmental setting.[57] During this phase, the first few cycles of swelling/deswelling with or without the stimulus cause microscopic changes and settling processes in the polymer network. These can result in an initially reduced repeatability and drift of the sensor response. However, through repeated cycling under the same conditions the repeatability is stabilized, which is the case for the studied composite.

The observed sample reaction to the water vapor flow purging can be explained as follows: Opening the inlet leads to a significant increase in water molecules in the chamber that compete with the acetone molecules in the composite fostering a removal of the organic solvent, and resulting in an increase in resistance. When the flow is turned off after 10 min and the sample left to settle again in the static humidity condition for another 10 min, more water molecules can be adsorbed on the surface, leading to a swelling change and an altered conductivity. Depending on the microstructure of the composite (2D or 3D, i.e., air- or freeze-dried, respectively), either one effect can be more pronounced. In the 2D case, the swelling/deswelling of the hydrogel component alters the spacing between the MXene parts, which ultimately causes the change of resistance (i.e. swollen hydrogel – larger spacing – higher resistance, and vice versa). For the 3D structure, the conductivity increasing effect appears to be more significant as reflected by the slight decrease in resistance.[58]

The 10 min time interval for purging has been chosen based on the ratio of chamber volume and amount of injected acetone and is deemed sufficient for complete removal of any organic solvent residue (also based on previous work[39]).

*3.2.4 Repeated chemiresistive analysis*

To evaluate the longer-term stability of the composite and the occurrence of degradation effects, multiple IDEs equipped with the MXene/PNIPAAm composite were repeatedly tested under the same conditions: cycling of acetone from 20 ppm to 100 ppm and back in water vapor atmosphere (approx. 160 min for one full cycle), storage for 1 week under cleanroom conditions (22 °C, 45 % RH, open box) and then retesting.

Please note that for this study, soft IDEs based on polyimide and with platinum metallization have been used (refer to Figure S6 in the Supporting Information for the IDE layout). The reason for this was that it proved difficult to obtain a stable adhesion of the composite to the rigid IDEs for prolonged amounts of time. The samples tended to fall off after synthesis, which was attributed to the mechanical mismatch at the interface between the glass-/ceramics-based IDEs and the comparatively soft MXene/PNIPAAm composite. By using a softer base material for the IDE, a larger number of intact samples could be obtained. This challenge only occurred for the composite material but not for the pure MXene.

Figure 6 depicts the IDE's delta resistance for each acetone condition and repeated testing. The delta resistance is calculated based on the sharp drop observed in the curves (Figure S9, Supporting Information). Please note that the first test, which was used to condition the polymer material in an acetone gas atmosphere to obtain a more stable output, is excluded.

From the diagram in Figure 6, it is evident that the sample exhibited an increased response to acetone in the third and fourth tests. However, in the fifth test, it dropped to a level similar to that of the second test. Nevertheless, compared to the second test where the material showed residual gas (indicated by an increased delta resistance at the same acetone concentration), the output in the fifth test demonstrated better stability and reproducibility.

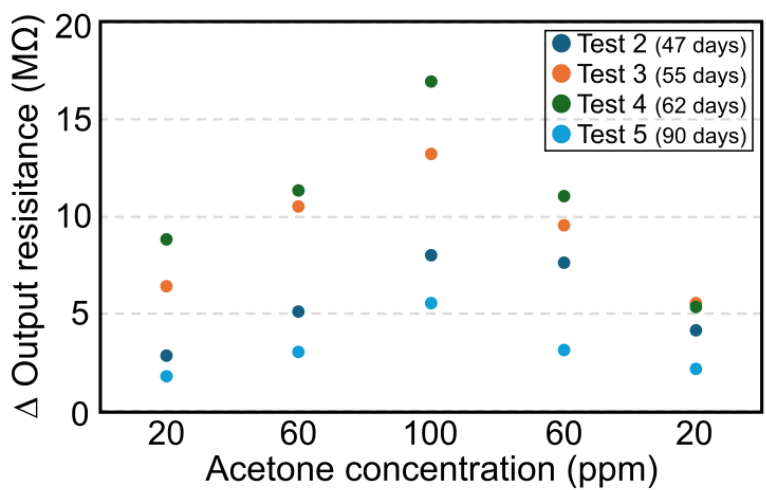

**Figure 6** Acetone response of repeated testing of the same MXene/PNIPAAm composite sample on IDE for five separate tests (details in Tables S2 and S3 in Supporting Information). The numbers in brackets denote the time after sample fabrication, i.e., the 'sample age'. The output resistance

change is calculated as the step height of the initial sharp drop as indicated by arrows in Figure 4d. Please note that the first test has been excluded as it is considered conditioning of the sample.

*3.2.5 Mechanical properties*

To investigate the influence of the MXene on the mechanical properties of the composite, corresponding bulk samples with varying MXene concentrations (ranging from 2 mg/mL to 25 mg/mL in the original suspension) were characterized by rheology (Figure 7). Pure PNIPAAm hydrogel with and without PEG was also included for comparison. Furthermore, all samples were tested in the watery swollen state.

Pure PNIPAAm without any additives or porogen shows the highest values for both, storage and loss moduli. By creating pores through the addition of PEG, the moduli are reduced and decline even further when MXene is added in low concentrations. This indicates a reduction of cross-linking strength.[59] However, with increasing MXene concentration (25 mg/mL and higher), the material becomes tougher again with the storage and loss moduli reaching similar or even slightly higher values than the PEG-modified PNIPAAm.

In all cases, the storage modulus is always larger than the loss modulus indicating the chemical cross-linking and solid-like behavior of the material. Furthermore, the storage moduli are almost independent on the oscillation frequency, while the loss moduli exhibit a more pronounced frequency dependence for all modifications (PEG alone and all MXene concentrations). This indicates an elastic deformation of the samples.

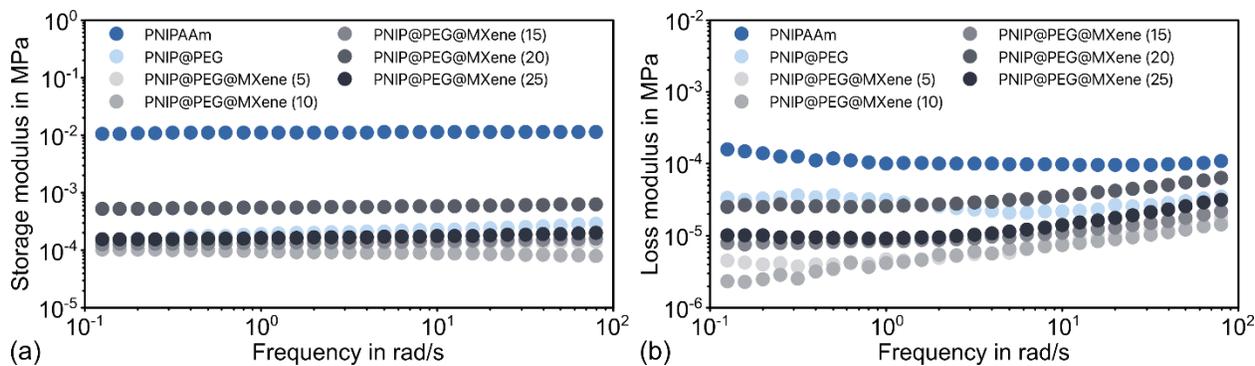

**Figure 7** Rheological test results of pure PNIPAAm and MXene/PNIPAAm composite samples with different MXene concentrations in the range of (5 – 25) mg/mL in the suspension (indicated by numbers next to composite in legend): a) storage and b) loss modulus in frequency scan mode. Angular frequency scan from 0.1 to 100 rad/s.

## 4. Discussion

### 4.1 Comparison of mechanical properties of pure MXene and MXene/PNIPAAm composite

As shown in Figure 7, the pure PNIPAAm hydrogel without added PEG for pore generation has the highest storage and loss moduli, and both values are substantially reduced when the porogen is used. The incorporation of low concentrations of MXene in the composite reduces the mechanical stability even further, since it affects the cross-linking density during polymerization. An increased MXene content yet improves the mechanical properties.

A similar phenomenon has been reported in other research, showing that relatively low and high MXene contents can produce transient structures dominated by MXene or by polymers, with higher yield stresses.[60] For structures containing MXene that lie in between, the lower filler content could result in a reduced crosslinking degree of the composite hydrogels.[59]

A further increase in MXene concentration will lead to the development of electrostatic interactions between the polar groups of the MXene nanosheets and the functional groups of the polymer chains, expanding the number of physical crosslinking points and ultimately improving the mechanical properties.[61] However, an excess of MXene in the polymer matrix leads to an increased viscosity of the precursor, making it challenging to fabricate intact samples of larger size. Furthermore, it can promote agglomeration and hinder the polymerization process. In our case, we found it impossible to create intact composite samples for MXene concentrations larger than 30 mg/mL. In summary, the MXene content can be used to tailor the mechanical properties of the MXene/PNIPPAm composite but needs to be carefully balanced with the precursor recipe and the amount of porogen.

### 4.2 The role of the porogen

The primary purpose of using a porogen with a hydrogel is the creation of an adjustable porous polymer network (bulk and surface). Our results for the MXene/PNIPAAm composite reveal that the porogen (here long-chain PEG) is also crucial for achieving a homogeneous dispersion of MXene in the hydrogel during polymerization. As depicted in Figure S7a (Supporting Information), the MXene/PNIPAAm composite synthesized by a simple mixture of monomer precursor and MXene suspension yields a heterogeneous structure after freeze-drying. The optical microscope image indicates a visible black-and-white pattern on the sample surface, while SEM images show dimpled surfaces of the composite. In addition, the sample exhibits an inhomogeneous combination of dense and porous structures in the cross-section. Previous work has shown that this indicates a high risk for pore collapse, i.e., low structural integrity.[39] The EDX mapping results (Figure S8a, Supporting Information) further validate this phenomenon: the partially collapsed structure of the plain MXene/PNIPAAm composite shows a pronounced accumulation of titanium.

A much more homogeneous MXene distribution within the composite is achieved by utilizing PEG during the polymerization. The combination of SEM and EDX analysis in Figures 4 and 5 evidence the homogeneous integration of MXene within the polymer matrix. Furthermore, the microstructure resembles a PEG-modified PNIPAAm sample without any filler (Figure S4), highlighting the predominance and scaffolding role of the polymer in this composite.[39] No individual MXene sheets can be identified in the SEM images and the EDX results (Figures 5 and S7b) clearly confirm a uniform titanium distribution.

These results highlight the importance of the interplay between hydrogel and porogen for the creation of a three-dimensional MXene arrangement. While the polymer assumes the scaffolding role, it is the spatial hindrance caused by the long-chain polymer molecules during composite synthesis that ultimately leads to a homogeneous MXene distribution. [62]

On one hand, the presence of PEG induces a phase separation of PNIPAAm during synthesis, leading to the formation of a macroporous hydrogel structure.[63] On the other hand, PEG creates steric hindrance in the crosslinking process, leaving behind inherent voids in the resulting polymer matrix after particle leaching.[63] This spatial constraint prevents the aggregation of MXene sheets into their usual stacked structure during the polymerization process.

Consequently, the use of a porogen (PEG in the presented case) is crucial for achieving a uniform dispersion of MXene flakes within the composite. This finding opens up the possibility of creating controlled porosity and spatial MXene distribution by the choice of the porogen (material type, molecular weight, amount) and the freezing temperature during synthesis. Moreover, with the presented synthesis process, the porous structure of the composite remains stable in gaseous environments with varying relative humidity as previously demonstrated[39] due to the scaffolding role of the polymer.

However, it is imperative to strike a balance between monomer concentration and porogen content to achieve an intact composite material. An increase of porogen may result in unsuccessful polymerization if the monomer concentration is too low in comparison.

## 4.3 Chemiresistive sensing analysis

In the chemiresistive sensing setup, a current is applied to the IDE, which passes through the sample placed on the electrode structure. The measured output resistance is proportional to the sample's conductivity that depends on the material properties as well as external influences such as adsorbed gas molecules. Consequently, the method can be used to detect gases and their concentration with a known sensing material, as well as for the characterization and comparison of different sensing materials (by using the same gas and concentration).

In the presented study, the chemiresistive analysis was used with the test organic gas acetone to evaluate the response and deduce material properties of the pure air-dried and freeze-dried MXene and the MXene/PNIPAAm composite.

In a comprehensive study, Pazniak et al. have investigated the conductivity mechanisms and responsiveness to organic gases of $Ti_3C_2T_x$ MXene in its pristine and partially oxidized state.[23] In its as-fabricated state, the material exhibits metal conductivity as a two-dimensional electron gas. Thus, the current in the IDE measurement passes through the interconnected MXene nanosheets with low resistance. It is furthermore hypothesized that the adsorption of gas molecules on the surface of the MXene flakes induces additional charge carriers whose presence is negligible due to the intrinsic metallic conductivity. However, they also create additional scattering centers, which cause an increase of the overall resistance, i.e., a reduction of conductivity. Consequently, irrespective of the gas species (reducing or oxidizing), pristine MXene always exhibits an increase of resistance upon gas adsorption.

The oxidation of the titanium component in the $Ti_3C_2T_x$ MXene substantially alters this conductivity behavior. $TiO_2$ itself is a wide-band semiconductor[64] and exhibits an insulting behavior for the chemiresistive sensing mechanism. During the transition of MXene from a pristine (metallic) to a fully oxidized (insulating) state, it exhibits a semiconducting behavior. The partial oxidation creates Schottky-type heterojunctions along and in between the MXene sheets. From the energy perspective, the continuous electron gas changes into the conductance/valence band structure with an energy barrier. However, this barrier is small enough that the additional charge carriers induced by adsorbed gas molecules increase the conductivity. In principle, they act in a similar way as donor or acceptor sites in a doped semiconductor.[56] As depicted in Figure 8b, an oxidizing gas shifts the Fermi energy towards the conduction band, while a reducing gas bends it closer to the valence band. Either way, due to the lowering of the barrier height in the Schottky-type heterojunctions, the conductivity of the partially oxidized MXene is increasing upon adsorption of gas molecules, i.e., the measured IDE output resistance is decreasing. However, the overall resistance for the current passing through the material is higher compared to the pristine MXene due to the numerous Schottky-heterojunctions.

As the oxidation progresses, it can be assumed that the energy barrier between valence and conduction band increases until it reaches a value where the change of the energy barrier height due to the injected charges from the gas adsorption becomes too small to have an effect on the charge carrier density in either band (Figure 8c). In this case, the material exhibits a very high resistance and becomes immeasurable with the chemiresistive technique.

Based on these considerations, the results from the chemiresistive analysis of the 2D (air-dried) and 3D (freeze-dried) pure MXene and its composite counterparts are discussed in the following.

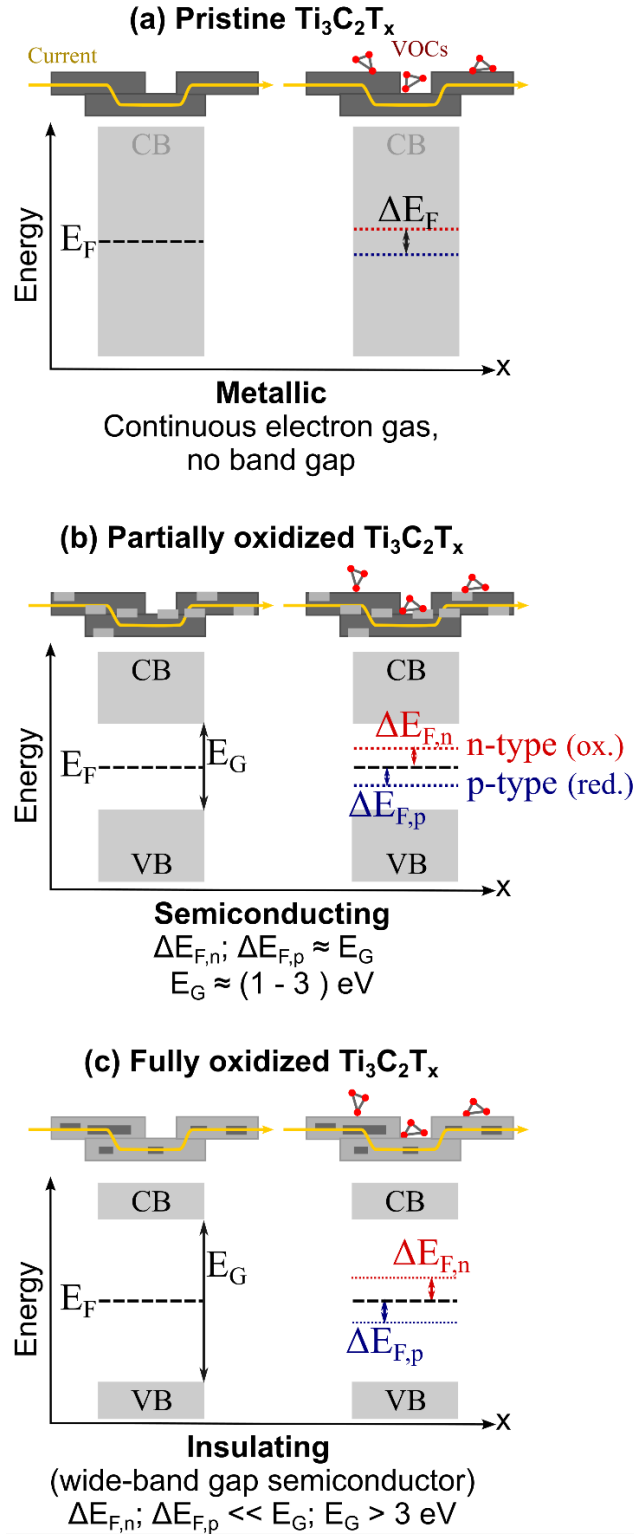

**Figure 8** Simplified principle of the conduction behavior of (a) pristine, (b) partially oxidized and (c) fully oxidized $Ti_3C_2T_x$ MXene. $E_F$, $\Delta E_{F,n}$ and $\Delta E_{F,p}$ denote the Fermi energy and its shift

towards the conduction (CB) or valence (VB) band in the presence of oxidizing or reducing gases, respectively. $E_G$ denotes the height of the energy barrier between both bands. In each diagram, the left and right parts depict the energy band structure without and with MXene, respectively. The sketches above the energy diagrams illustrate the current passing through the linked MXene sheets, the VOC adsorption and the oxidation state (dark grey: pristine MXene, light grey: $TiO_2$).

*4.3.1 Pure MXene*

According to the results of the chemiresistive response to varying acetone concentrations depicted in Figure 2c, pure 2D MXene exhibits an increase in resistance independent on the acetone concentration with an overall resistance of a few hundred Ohms. This is consistent with a metallic conductivity behavior and the findings of other publications.[3,53,65]

In contrast, the results for the 3D MXene structure depicted in Figure 2d indicate a semiconducting response since the resistance decreases in the presence of acetone, while the overall resistance is much larger (Megaohms) compared to the 2D sample. It can be assumed that this change of conductivity behavior is due to the highly porous structure of the material visible in the SEM images. The substantially increased surface area compared to 2D MXene fosters oxidation and, thus, the transition into the semiconducting state. The progression of the oxidation is furthermore evident in the repeated testing of the sample with the same acetone concentrations (Figure 2d), where the resistance kept increasing over six weeks until it became immeasurable (insulating state). In addition, the results of the four-week sample test show a clear dependence of the resistance decrease on the acetone concentration for the range of (20 – 80) ppm. For 100 ppm the decrease is reduced which is attributed to a saturation effect of the material.

It needs to be noted here that the chemiresistive analysis of all pure MXene samples has been conducted with a dried nitrogen background instead of air. It is thus expected that the material would have degraded and lost its responsiveness even earlier if a humid background similar to that of the composite had been used.

The oxidation of the titanium in the MXene occurs randomly and rather uncontrolled. However, our findings indicate that one potential way to exert at least some influence on the progression of oxidation is by controlling the porosity of the surface and bulk parts of the material, as well as the pore size by the freezing temperature and drying method.

*4.3.2 MXene/PNIPAAm composite*

The chemiresistive response of the composite material is more complex as both, the MXene and the PNIPAAm hydrogel, respond to the acetone gas, and, in particular the hydrogel, also to the humid air. It needs to be noted that the humidity is necessary to create a measurable output resistance. With a dried nitrogen background as in the case of pure MXene, no output signal could be obtained. It is therefore assumed that the polymer at least partially passivates the MXene and reduces the contact between individual sheets. In its fully dried state, the composite is non-conductive but the presence of water molecules enables current pathways through the material.

Comparing the chemiresistive results of the 2D (air-dried) and 3D (freeze-dried) composite depicted in Figure 4c and 4d, it is clearly evident that the 2D sample does not show any detectable resistance change to varying acetone concentrations. It only undergoes a conditioning process in the first two cycles where the resistance is slightly increasing, likely due to the relaxation and rearrangement of the polymer chains in response to the higher relative humidity (RH = 100 %) in the test chamber compared to the cleanroom storage conditions (RH = 45 %). This behavior can be attributed to the non-porous and skin-like sample surface visible in the SEM images. This likely prevents water and gas molecules from penetrating the material body and initiating a response.

In contrast, the very open and interconnected porous structure obtained by freeze-drying creates a very large surface area for the interaction with water molecules from the humid air background as well as the adsorption of acetone gas molecules. The observed response to the presence of acetone (Figure 4d) can be attributed to two effects: (i) the interaction of MXene with acetone, and (ii) the physical swelling response of the hydrogel that in itself would cause an increase in resistance. These two effects exhibit different time constants. While the charge transfer interaction between MXene and the gas molecules happens very fast, the hydrogel swelling is a diffusion-based process that relies on the analyte being adsorbed on the surface and subsequently reaching the inner part of the polymer network. This causes a comparatively slow response until the hydrogel reaches a swelling equilibrium. Considering the curve shape of the acetone response depicted in Figure 4d, it can be assumed that the initial sharp drop in resistance is caused by effect (i), since it is consistent with the conductivity behavior of partially oxidized MXene.

Once the hydrogel starts to notably swell, the effect is partly counteracted by the increasing distance between the MXene flakes embedded in the polymer. This likely leads to the second part of the curve where the resistance keeps decreasing but with a substantially reduced slope compared to the initial drop. Whether the step height of the initial drop or the slope of the slowly decreasing resistance is more informative in terms of a sensing application for acetone needs to be evaluated in further studies.

The overall resistance of the 3D composite sample is comparable to the pure 3D MXene, indicating that the base conductivity is determined by the partially oxidized MXene and not the hydrogel or the presence of water molecules.

## 4.4 Degradation stability

The oxidant responsible for the MXene transition from metallic conductivity to semiconducting and subsequently insulating is the oxygen from air and water molecules.[66] During sample storage under cleanroom conditions, this non-specific oxidation progresses over time as evidenced by the changing chemiresistive response.

Furthermore, the specific experimental conditions (acetone exposure, high humidity in case of the composite material) during the chemiresistive measurement itself may also contribute to the degradation. However, the duration of one full-cycle experiment that covers different acetone concentrations is only 160 mins, which is negligible compared to the constant exposure to humid air during weeks of sample storage.

By comparing the degradation progression of the 3D pure MXene and composite sample, the role of the hydrogel can be deduced. In both cases, the highly porous three-dimensional microstructure fosters the initiation of the oxidation processes right after sample fabrication. For 3D pure MXene, the oxidation progresses continuously until the output resistance becomes immeasurably high within 6 weeks after fabrication. In contrast, repeated chemiresistive cycling of the composite (Figure 6) shows that the resistance change first becomes larger with retesting (tests 3 and 4) and then starts to decrease again (test 5) over a period of three months (almost 13 weeks). This indicates a passivation effect of the PNIPAAm hydrogel and a substantially prolonged oxidation stability compared to pure 3D MXene. However, 15 months after fabrication, the composite sample is fully oxidized and immeasurable as well. This is corroborated by the EDX results for both, the pure 3D MXene and composite samples, which indicate a similar doubling of the oxygen content (Figures 3 and 5).

These initial investigations indicate the potential of the studied MXene/PNIPAAm composite with respect to VOC sensing, but also for an advanced future material design where the properties of both components are combined to create new functionalities. Our study shows that the combination of the developed porosity engineering techniques for polymer materials, such as the use of a porogen, variation of drying method and freezing temperature, can substantially alter the microstructural properties and spatial distribution of MXene in a hydrogel composite, and consequently its degradation stability and conductivity behavior. The creation of electronic states is affected by the interplay between polymer and MXene, which furthermore reflects on the potential usability of the composite as a gas sensing material.

## 5. Conclusion

This work presents a fundamental investigation of the properties of a novel dry MXene/PNIPAAm composite, with a particular focus on its potential for an application in volatile organic compound (VOC) detection in gaseous environments using chemiresistive transduction. Therefore, pure MXene and composite samples featuring either a compact and dense or a highly porous three-dimensional microstructure, have been fabricated and characterized. These investigations reveal advanced properties with respect to MXene degradation stability and control over the three-dimensional structural arrangement.

With regard to electronic properties, the results show that the tailored porosity of pure MXene, namely from a densely packed to a fully three-dimensional highly porous microstructure, allows for alteration from metallic to semiconducting behavior in response to a VOC interaction in a chemiresistive sensing configuration. This is due to the oxidation of the titanium in MXene being dependent on, and therefore adjustable by, the porosity. Furthermore, the creation of partially oxidized MXene is simply achieved through the tailoring of the 3D structure without the need for the commonly used additional heating.

The oxidation of pure 3D MXene under ambient conditions occurs rather rapidly and uncontrolled, leading to the loss of conductivity and responsiveness to gas interactions. In contrast, the composite exhibits a significantly slowed degradation behavior. It maintained its responsiveness and semiconducting state for at least 12 weeks (pure 3D MXene: 6 weeks) under repeated testing in the same variations in acetone levels with storage under ambient cleanroom conditions in between tests. This is attributed to a passivation influence of and interplay with the polymer.

The underlying mechanisms of this discovery need to be studied further. So far, two major implications can be derived: (i) The composite significantly reduces the speed of the oxidation-induced degradation of the MXene, thereby stabilizing the electronic state and interaction with gas species. (ii) The porosity characteristics (e.g., pore distribution, size and interconnection) enable control and adjustment of the composite's electronic state and sensitivity to gas interactions. On the one hand, a larger number of pores can increase the resistance. On the other hand, it also creates a larger surface area for the oxidation of the titanium and consequently, the formation of Schottky-heterojunctions that increase the responsiveness to gas adsorption.

The hydrogel provides a 3D scaffold into which the MXene is fully integrated without the usual stacking of sheets. The microstructure, spatial MXene arrangement, and porosity of the composite can be tailored by the choice of the porogen, the drying method (air- or freeze-drying) and the freeze-drying temperature. Thus, the application of the porosity engineering techniques developed for the polymer enable fundamentally novel avenues for tuning and tailoring MXene properties with the integration into a composite material.

From the perspective of a sensing application for VOC detection, the control of the interplay between the polymer and MXene through the microstructure can be employed to adjust the analyte sensitivity from almost non-responsive (for a dense and compact microstructure) to strongly responsive (for 3D porous). In the latter case, the distinctly different interaction mechanisms of charge transfer of the MXene and mechanical swelling/deswelling of the hydrogel with the test analyte acetone are reflected in a very distinct resistance change pattern. In the context of gas sensing, this enables: (i) a high sensitivity to acetone concentrations as low as 20 ppm, (ii) a stable and reproducible response, and (iii) the potential for very fast tracking of changes in analyte concentration (within a timeframe of less than 2 min).

Moreover, the described techniques for the fabrication of a tailored microstructure of the composite are of general nature and not specific for the PNIPAAm hydrogel used in this study. Hence, they can be applied to any other type of polymer and its integration with MXene, thereby enabling the development of sensing applications for a wide range of analytes due to the versatility of the polymer.

From the overall perspective of advancing material science, the presented study demonstrates the creation of a composite by merging the well-known class of polymers with the novel group of MXenes. By employing techniques originally developed for the porosity engineering of hydrogels, the resulting material features a complex interplay between both constituents that offers great potential for addressing some of the remaining major challenges associated with MXene use to date, namely oxidation-induced material degradation, control over electronic state and creation of a fully three-dimensional microstructure.

ASSOCIATED CONTENT

**Supporting Information**.

The following files are available free of charge.

Supporting Information containing: 1. Details of sample fabrication and characterization; 2. Supporting figures; 3. Supporting tables; 4. References (file type .pdf).

AUTHOR INFORMATION

**Corresponding Author**

*Julia Körner – Department of Electrical Engineering and Computer Science, Leibniz University Hannover, 30167 Hannover, Germany; Email: koerner@geml.uni-hannover.de

**Author Contributions**

The manuscript was written through contributions of all authors. All authors have given approval to the final version of the manuscript. S.W.: Conceptualization, data curation, methodology, investigation, formal analysis, visualization, and writing - original draft. C.J.: Investigation, and writing - review and editing. G.G.: Resources, supervision, writing - review and editing, project administration, and funding acquisition. J.K.: Conceptualization, validation, supervision, visualization, writing - review and editing, and project administration.

**Funding Sources**

This work was supported by the German Research Foundation (Deutsche Forschungsgemeinschaft, DFG) in the framework of the Research Training Group "Hydrogel-based microsystems" (GRK 1865).

ACKNOWLEDGMENT

The authors would like to thank Yuhao Shi, Stefan Schreiber and Alice Mieting for helpful discussions, Daniela Franke and Margarita Günther for theoretical and practical guidance, Maik Müller for assistance in SEM operation, Benozir Ahmed for fabrication of the soft IDEs and Christopher Reiche for critical review of the final manuscript.

**TOC graphic**

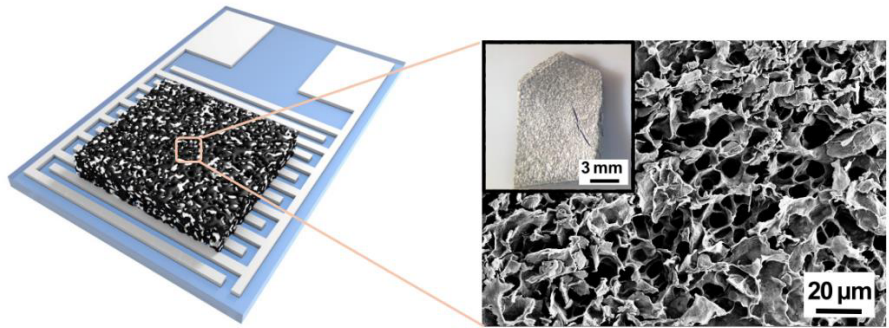

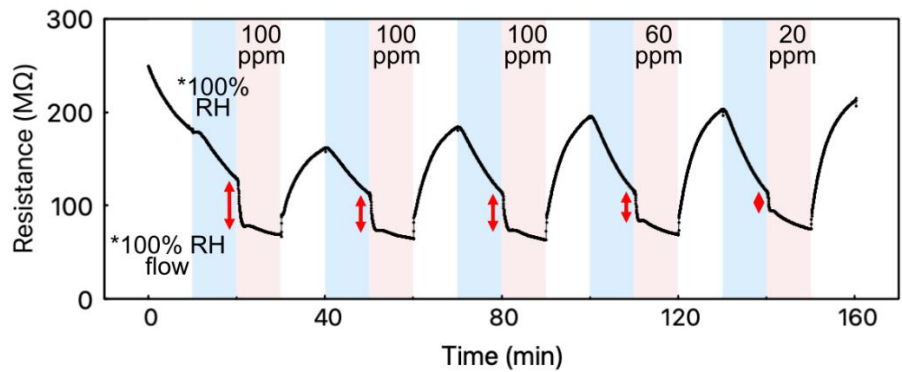

# Supporting Information

**A 3D porous MXene/PNIPAAm hydrogel composite with advanced degradation stability and control of electronic properties in air**

*Sitao Wang[1], Chen Jiao[2], Gerald Gerlach[1], and Julia Körner[3,*]*

[1]Institute of Solid-State Electronics, Dresden University of Technology, 01062 Dresden, Germany

[2]Leibniz Institute of Polymer Research Dresden, 01069 Dresden, Germany

[3]Department of Electrical Engineering and Computer Science, Leibniz University Hannover, 30167 Hannover, Germany

*E-mail: koerner@geml.uni-hannover.de

## Content

1. Details of sample fabrication and characterization
2. Supporting figures
3. Supporting tables
4. References

## 1. Details of sample fabrication and characterization

### 1.1 Sample fabrication

*Fabrication of bulk pure MXene structure:* 3D MXene structures were fabricated by freezing-induced preassembly and a subsequent treatment in protic acids.[S1,S2] By shifting their surface zeta potential in protic acids like hydrochloric acid (HCl) and sulfuric acid ($H_2SO_4$), negatively charged MXene nanosheets cross-link and form a 3D structure due to the electrostatic interaction. These obtained 3D scaffolds are also commonly denoted as MXene hydrogels. In the presented case, $Ti_3C_2T_x$ powder was purchased from Nanoplexus Ltd. (England, UK) and used directly without further purification. To obtain a homogeneous MXene suspension, 30 mg powder was dispersed in 1 mL deionized water and sonicated in an ice bath for 3 h. Then the $Ti_3C_2T_x$ dispersion was poured into glass molds with a 1200 μm thick PTFE spacer. The mold edges were secured with binder clips (19 mm) to prevent leaking of liquid. For ice templating, the molds were either placed in liquid nitrogen (-196 °C) for 5 min and subsequently transferred to a refrigerator (-20 °C) or kept directly in the refrigerator for 24 h. Afterwards, the frozen sample was put into 12 M HCl (37%, VWR, Germany) solution and stored for another 24 h.

The acid solution was then replaced with deionized water for 3 days with daily solution change until the acid had been completely removed. Ultimately, to preserve the porous structure, the samples were freeze-dried with liquid nitrogen (-196°C).[S3,S4]

*Fabrication of pure MXene structure on IDE:* Due to the fragile nature of the porous pure MXene structure, using a mold system with a thinner spacer is not feasible. Therefore, a simple drop-casting method was employed (Figure S1a). In this method, a hemispherical droplet was manually pipetted onto the IDE surface (glass substrate), consisting of 8 μL of MXene suspension. Subsequently, the entire IDE with the deposited material on top was either air- or freeze-dried as described below.

*Fabrication of bulk MXene/PNIPAAm composite with PEG porogen:* The MXene/PNIPAAm composites were prepared by simply mixing the hydrogel with the MXene suspension. PEG was employed as a pore-forming agent, which was removed after polymerization by washing in deionized water for 7 days with daily liquid exchange. Specifically, 1 mL of MXene suspension at a concentration of 30 mg/mL was mixed with 1 mL of NIPAAm precursor containing 1 g of PEG.

To prepare the hydrogel precursor, NIPAAm (4.42 mmol, 0.5 g) and MBA (0.21 mmol, 0.033 g) were fully dissolved in 3 mL of deionized water and degassed with nitrogen for 5 min. Then, TEMED (0.05 mmol, 7.5 μL) and APS (0.07 mmol, 0.0167 g) were added, followed by 1g of PEG as porogen. After 3 days of polymerization in a cleanroom environment and leaching of PEG in deionized water for another 3 days, the sample was either freeze-dried by freezing in liquid nitrogen (-196 °C) or air-dried. For more details on the utilization of PEG, please refer to our previous work.[S4]

*Fabrication of MXene/PNIPAAm composite with PEG porogen on IDE:* Due to the mechanical flexibility of polymers, it is possible to achieve precise control over the composite thickness and shape with a mold in case of the MXene/PNIPAAm on IDEs (in contrast to pure MXene). The IDE was initially placed within a mold system, where a Teflon spacer of a defined thickness (100 μm) was positioned between two glass slides (Figure S1b). The precursor solution was then injected into the space between one of the glass sheets and the IDE surface, resulting in the successful fabrication of a thin layer of MXene/PNIPAAm composite on top of the IDE. Following a 3-day polymerization period, the desired material shape (full coverage of the IDE electrode structure) was obtained by carefully cutting it with a blade. The leaching of PEG, as previously described, was continued for another 3 days by submerging of the IDE sample in deionized water with daily liquid exchange. Finally, the sample was either freeze- or air-dried.

*Freeze-drying of fabricated hydrogel-containing samples:* After being fully rinsed in deionized water (resulting in a swollen state), the samples were transferred into 50 mL glass vials compatible with the connectors of the freeze-dryer. There were two options for freezing: i) immersion of the samples in liquid nitrogen (-196 °C) for 10 min, followed by transfer to the freeze-dryer, which maintained reduced pressure for the gradual sublimation of ice; or ii) placement of the samples in a freezer (-20 °C) for 24 h before moving them to the freeze-dryer. In both cases, all samples were connected to the drying equipment for a continuous 24 h period to ensure the complete sublimation of ice crystals.

## 1.2 Materials and chemicals

N-isopropylacrylamide (NIPAAm), N,N'-methylenebis(acrylamide) (MBA), ammonium persulfate (APS), N,N,N',N'-tetramethyethy-lenediamine (TEMED), lithium phenyl-2,4,6-trimethylbenzoylphosphinate (LAP) were purchased from Sigma-Aldrich (Germany) and were used as received. Poly(ethylene glycol) (PEG) with molecular weight of 10000 was purchased from FLUKA (Switzerland) and used as supplied. MXene ($Ti_3C_2T_x$) powder was purchased from Nanoplexus Ltd. (United Kingdom) and used directly without further purification. 37% hydrochloric acid (HCl) solution was obtained from VWR (Germany).

## 1.3 Sample treatment and characterization

*Air-drying of fabricated hydrogel-containing:* After polymerization and rinsing for 7 days in DI water, the swollen sample was removed from the water and left to dry in open air under cleanroom conditions (22 °C, 45% RH) for 3 days. Based on previous studies, this ensures complete removal of liquid.[S3,S4]

*Sample characterization:* Optical microscope images were taken with a stereo microscope (Leica MZ6, Switzerland) with lighting (SHOTT KL 750, Germany). The detailed surface morphology and internal structure of the materials were characterized by using a Zeiss Supra 40 VPF scanning electron microscope (Carl Zeiss GmbH, Germany). For hydrogel-based materials, the samples had to be dehydrated and coated with 10 nm of gold to avoid charging effects due to the low electrical conductivity of the polymers. Gold deposition was performed by sputter coating (SC7620 Mini Sputter Coater, Polaron, Germany). For 3D MXene structures, no gold deposition was required. Additionally, EDX (EDAX, ELECT PLUS, AMETEK, USA) was employed to study the elemental distribution in the samples. Samples were prepared in the same manner as for SEM. Hydrogel-based bulk samples were assessed by rheology using a strain-controlled rheometer (ARES-G2, TA Instruments, USA) with a parallel plate geometry of 25 mm in diameter. With a fixed oscillatory strain of 1%, a dynamic frequency sweep in the range from 0.1 to 100 rad/s was conducted at 22 °C.

*Evaluation of gas sensing performance:* To investigate the chemiresistive response of the sensing materials to varying levels of acetone, the chip was placed in a sealed chamber where the desired environmental conditions could be produced by injecting a defined volume of liquid organic solvent (see Figure S2 for calculation). The acetone concentration in the chamber was adjusted from 20 to 100 ppm, with steps of 20 ppm in between.

Two testing procedures were employed:

i) For pure MXene samples: The sample was placed in the chamber followed by purging with a dried nitrogen flow. Subsequently, the gas flow was turned off, absolute acetone injected, and the chamber sealed. Following exposure to the organic environment, the chamber was again purged with the dried nitrogen flow before continuing to the next gas atmosphere (acetone concentration). The duration of each condition was 10 min.

ii) For hydrogel-containing samples: This second procedure was applied for all hydrogel-based samples to ensure sufficient hydration of the polymer component. A liquid water reservoir was consistently present in the chamber throughout the test to generate water vapor. The sample was initially placed in the chamber, which was then rinsed with a humid gas flow produced by the bubbler. In the subsequent step, the gas flow was stopped, absolute acetone injected, and the chamber sealed. After exposure to the organic gas, the chamber was purged again with the same humid gas flow before moving on to the next acetone condition. In this study, all gas conditioning intervals, including exposure to the organic atmosphere and purging of the chamber, were set to 10 min.

In both testing cases, the IDE resistance was measured by a digital multimeter (Fluke 45 Dual Display Multimeter, USA) and the output results recorded with a self-programmed script (one data point every 2 s). The gas flow rate was measured and controlled by a mass flow controller (Bronkhorst EL-flow with a nominal flow rate of 5 l/min).

## 2. Supporting figures

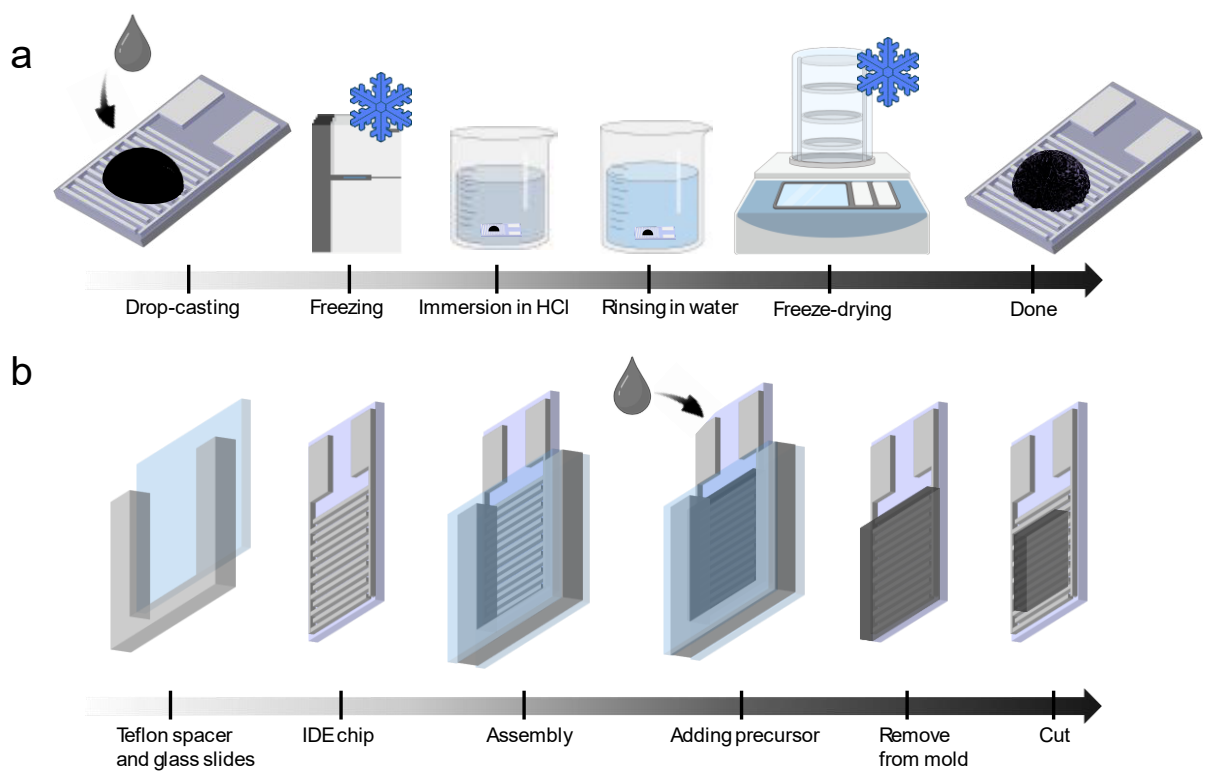

**Figure S1** Fabrication process of (a) pure MXene and b) MXene/PNIPAAm composite samples on IDE.

$$\frac{c}{\mathrm{ppm}} = \frac{\frac{24.22}{\frac{\mathrm{L}}{\mathrm{mol}}} \cdot \frac{T}{\mathrm{K}} \cdot \frac{\rho_\mathrm{f}}{\frac{\mathrm{g}}{\mathrm{mL}}} \cdot \frac{V_\mathrm{f}}{\mathrm{mL}}}{\frac{273}{\mathrm{K}} \cdot \frac{M}{\frac{\mathrm{g}}{\mathrm{mol}}} \cdot \frac{V}{\mathrm{L}}} \cdot 10^3$$

24.22 L mol$^{-1}$……ideal gas volume at 22 °C (295 K) and one standard atmosphere pressure (101.325 kPa; 760 mm Hg)

273 K……equals 0 °C in Kelvin scale

$c$……gaseous analyte concentration (ppm)

$T$……working temperature in K (295 K in cleanroom)

$\rho_\mathrm{f}$……density of the liquid organic solvent (g mL$^{-1}$)

$V_\mathrm{f}$……volume of liquid solvent (mL)

$M$……the molecular weight of the analyte (g mol$^{-1}$)

$V$……the volume of the testing chamber in liter

| Acetone concentration | 20 ppm | 60 ppm | 100 ppm |
|---|---|---|---|
| Injected liquid acetone | 0.02 mL | 0.06 mL | 0.10 mL |

**Figure S2** Determination of the amount of liquid acetone for reaching predefined concentrations: a) Calculation formula and corresponding parameters, b) volume of the injected liquid acetone. For details refer to [S3].

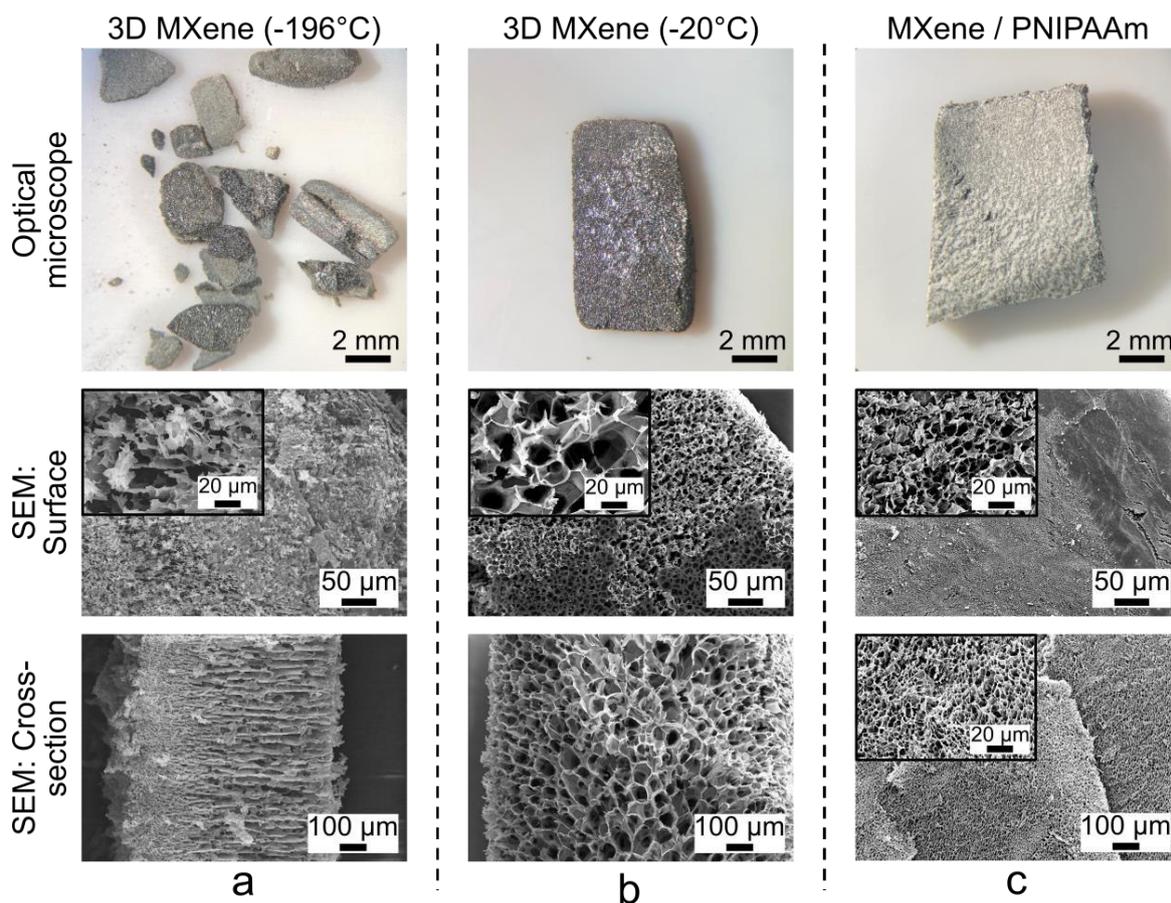

**Figure S3** Optical microscope and SEM images (surface and cross section) of freeze-dried samples: a) pure MXene fabricated by ice-templating at -196 °C. b) Pure MXene fabricated by ice-templating at -20 °C. c) MXene/PNIPAAm composite fabricated at room temperature. All samples were freeze-dried using liquid nitrogen.

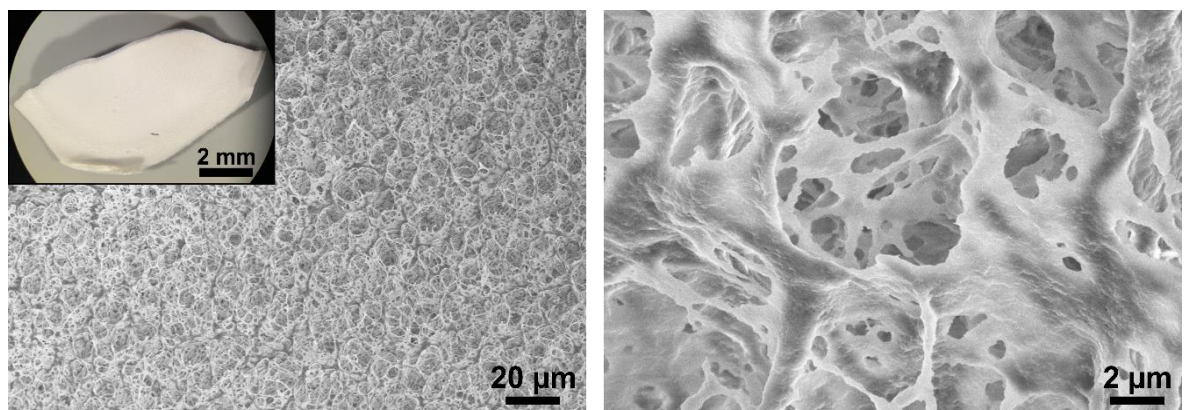

**Figure S4** Optical microscope and SEM images (left: surface, right: cross section) of freeze-dried (by liquid nitrogen) pure PNIPAAm hydrogel with PEG as porogen that was washed out after polymerization.[S4]

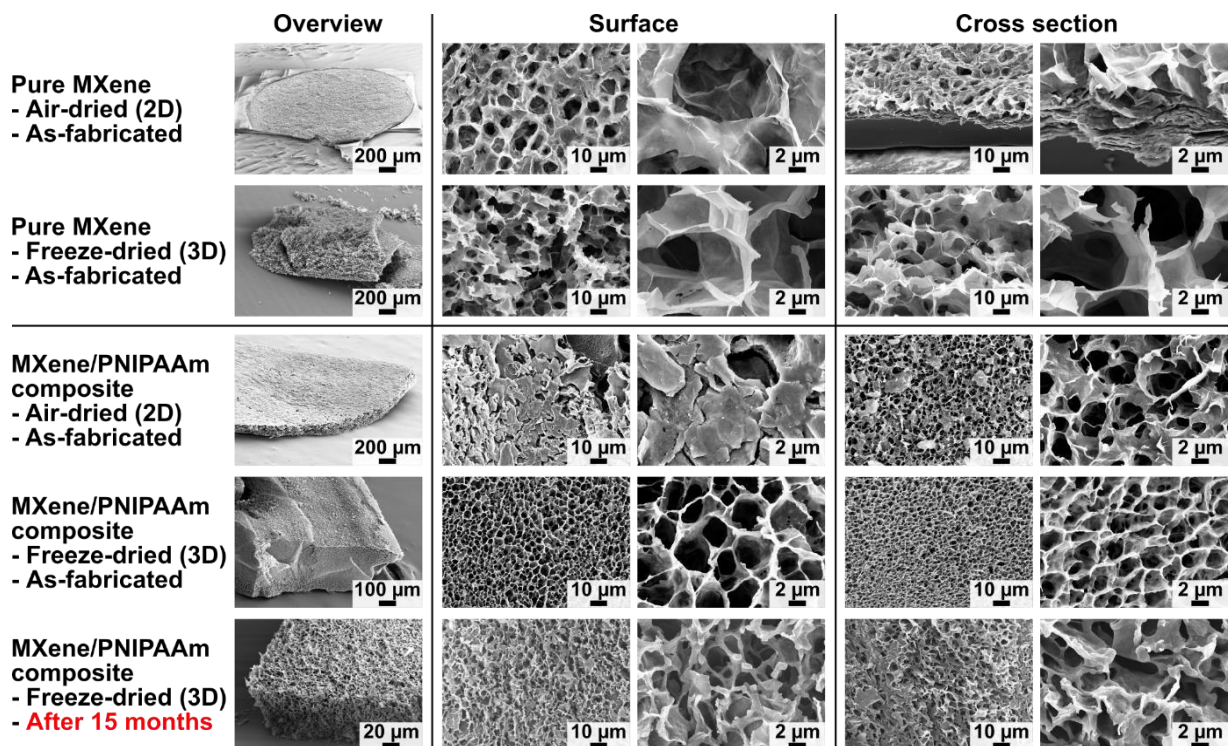

**Figure S5** Detailed SEM images of all on-chip sample types (air-dried, freeze-dried, pure MXene, MXene/PNIPAAm composite). For the composite, as-fabricated and appearance after 15 months of repeated testing and storage are shown. An overview (left) and a larger magnification (right) are depicted in the *surface* and *cross section* columns.

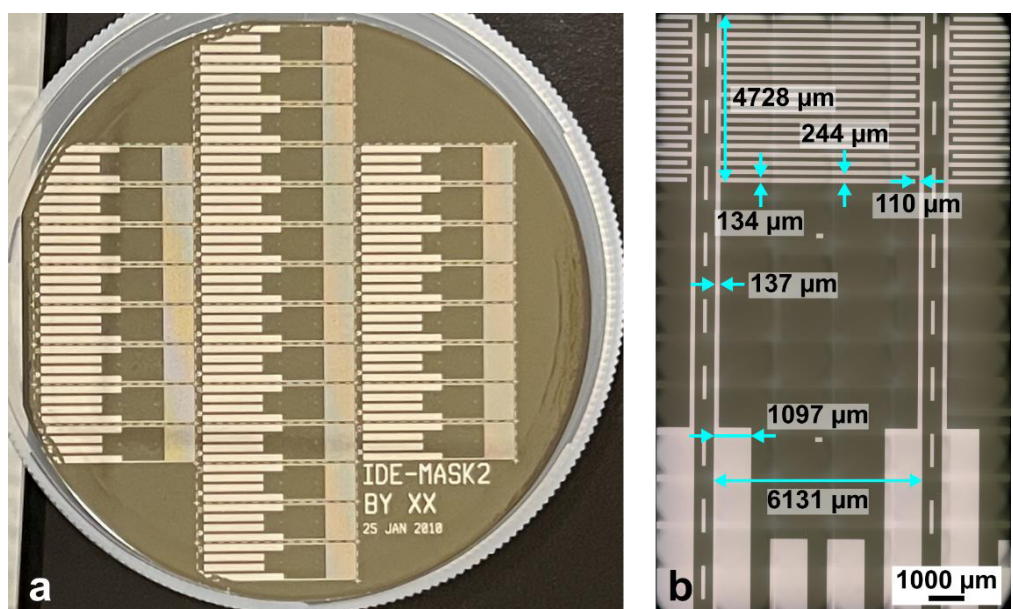

**Figure S6 a)** Platinum IDEs on polyimide on a 4" silicon carrier wafer (photographic image). **b)** Dimensions of one IDE electrode. IDEs were fabricated by Benozir Ahmed from the Department of Electrical and Computer Engineering at the University of Utah, USA.

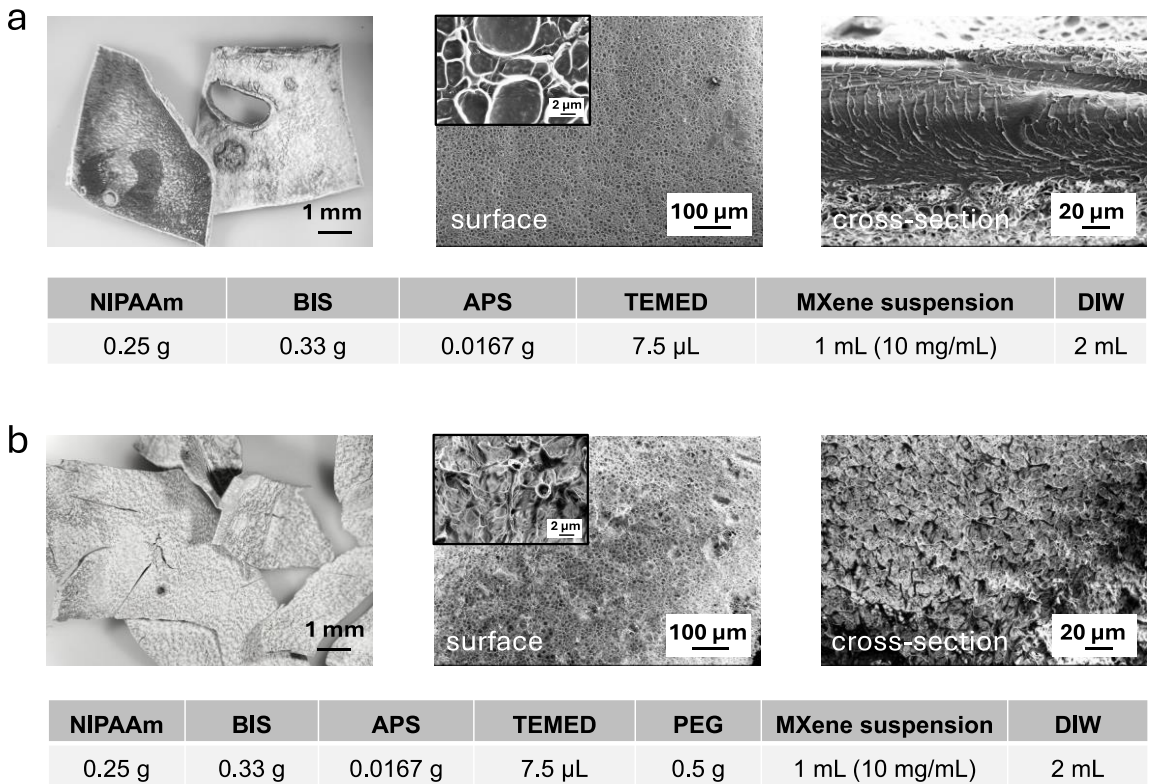

**Figure S7** Synthesis recipes and optical as well as SEM images (surface, cross-section) of MXene/PNIPAAm composites. a) Plain MXene/PNIPAAm fabricated by mixing of NIPAAm precursor and MXene suspension. b) PEG-modified MXene/PNIPAAm with PEG as porogen that is washed out after polymerization.

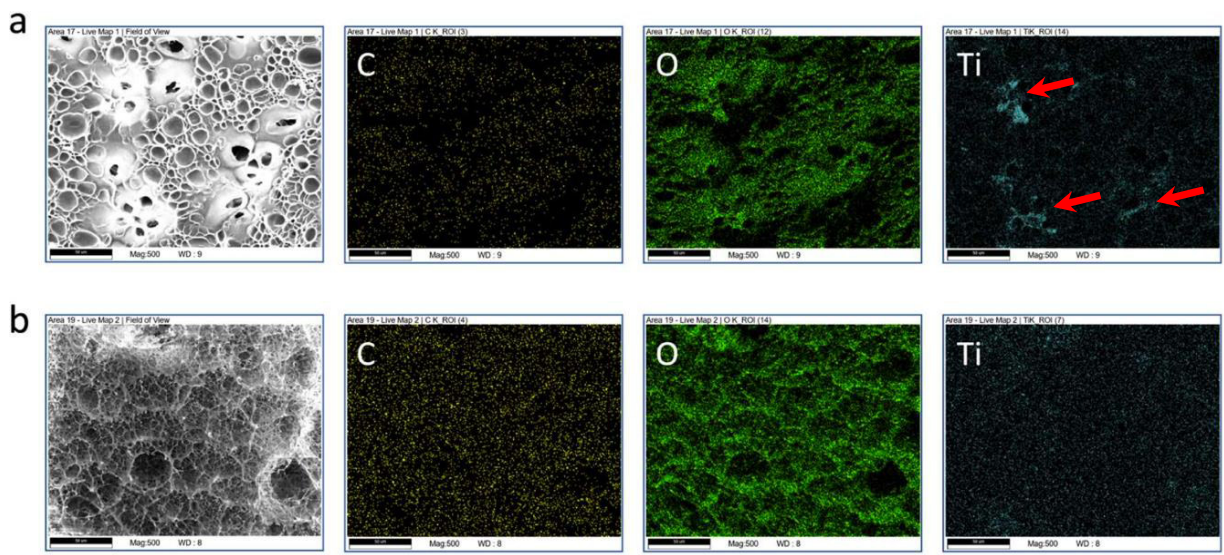

**Figure S8** Energy-dispersive X-ray (EDX) spectroscopy mapping results of MXene/PNIPAAm composites: a) Unmodified (without porogen) and b) PEG-modified. In both cases, the samples were fabricated from a MXene suspension with a concentration of 10 mg/mL. The red arrows in the Ti mapping in subfigure (a) indicate titanium aggregation. This effect only occurs without the use of a porogen.

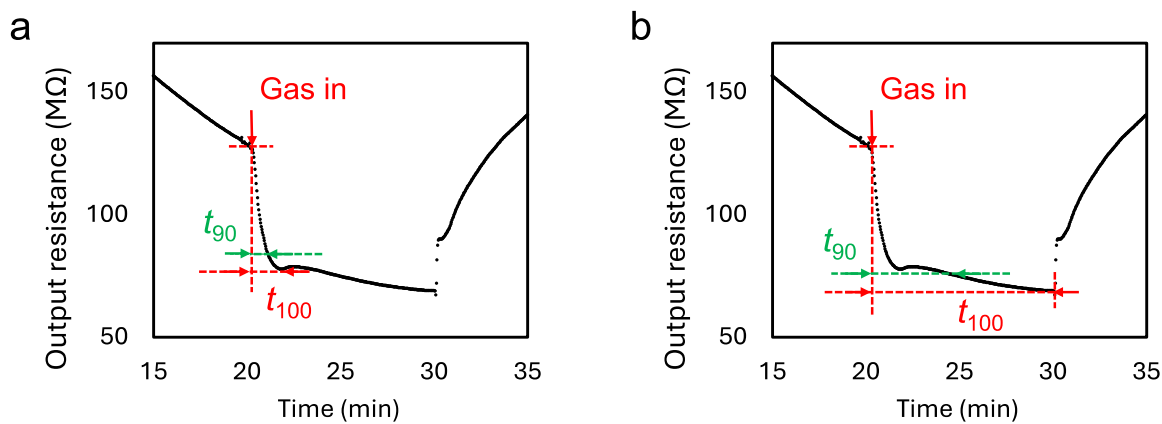

**Figure S9** Determination methods of the response time $t_{90}$. a) Calculation based on step height of the initial sharp drop. b) Calculation based on the overall resistance decrease (90 % of steady-state value). The first method has been used for the determination of the response times listed in table S2.

## 3. Supporting tables

**Table S1** Calculated response time $t_{90}$ (min) and magnitude of initial IDE resistance drop (in MΩ) according to Figure 4 (main text) and Figure S9

| Acetone concentration | 100 ppm (1st) | 100 ppm (2nd) | 100 ppm (3rd) | 60 ppm | 20 ppm |
|---|---|---|---|---|---|
| Response time (acc. to S9a) | 1.18 | 1.13 | 1.3 | 0.86 | 0.67 |
| Response time (acc. to S9b) | 5.84 | 5.5 | 5.71 | 6.58 | 7.24 |
| Magnitude of initial drop (S9a) | 50.3 | 39.6 | 41.1 | 31.8 | 20.7 |
| Magnitude of initial drop (S9b) | 60.8 | 48.1 | 50.7 | 46.5 | 40.6 |

**Table S2** Timeline for repeated tests of the porous MXene/PNIPAAm composite on IDE in Figure 6 (main text) spanning 3 months

| Test date | Test sequence | Gaseous conditions |
|---|---|---|
| 2024/01/08 | 1st | 100/100/100/60/20 ppm (conditioning) |
| 2024/02/23 | 2nd | 20/60/100/60/20 ppm |
| 2024/03/02 | 3rd | 20/60/100/60/20 ppm |
| 2024/03/09 | 4th | 20/60/100/60/20 ppm |
| 2024/04/06 | 5th | 20/60/100/60/20 ppm |

**Table S3** All test data of the MXene/PNIPAAm on-chip, freeze-dried composite sample (Figure 6 in main text) for the three-month repeated testing (timeline table S3). The resistance changes ΔR (in MΩ) have been calculated based on the initial drop after the introduction of acetone. The first test is excluded as it followed a different acetone concentration cycling for conditioning.

| Test / Acetone conc. | 20 ppm | 60 ppm | 100 ppm | 60 ppm | 20 ppm |
|---|---|---|---|---|---|
| 2$^{nd}$ test | 2.9 | 5.2 | 8.1 | 7.7 | 4.2 |
| 3$^{rd}$ test | 6.5 | 10.6 | 13.3 | 9.6 | 5.6 |
| 4$^{th}$ test | 8.9 | 11.4 | 17.0 | 11.1 | 5.4 |
| 5$^{th}$ test | 1.9 | 3.1 | 5.6 | 3.2 | 2.2 |